\documentclass[11pt,a4paper]{article}
\usepackage{jheppub}
\usepackage{epstopdf}
\usepackage{amssymb}

\usepackage{inputenc}
\usepackage{xspace}
\usepackage{amsfonts}
\usepackage{url}

\newcommand{\bea}{\begin{eqnarray}}
\newcommand{\eea}{\end{eqnarray}}
\newcommand{\bean}{\begin{eqnarray*}}
\newcommand{\eean}{\end{eqnarray*}}
\newcommand{\nn}{\nonumber\\}

\def\Tr{\mathop{\rm Tr}}

\def\Label#1{\label{#1}%
  \smash{\hbox to0pt{\raise1ex\hbox{\tiny[#1]}\hss}}}

\def\bse{\begin{subequations}}
\def\ese{\end{subequations}}

\def\Sl{\sum\limits}

\preprint{LU-TP 14-42}

\allowdisplaybreaks

\title{Note on off-shell relations in nonlinear sigma model}

\author[a]{Gang Chen}
\author[b,1,2]{Yi-Jian Du%
\note{Corresponding author.}
\note{On leave from Center for Field Theory and Particle Physics, Department of Physics, Fudan University, China.}}
\author[a]{Shuyi Li}
\author[a]{Hanqing Liu}
\affiliation[a] {Department of Physics, Nanjing University,\\
 22 Hankou Road, Nanjing 210093, China}
\affiliation[b]{Department of Astronomy and Theoretical Physics, Lund University,\\
  S\"olvegatan 14A, 223\,62~Lund, Sweden}

\emailAdd{gang.chern@gmail.com}
\emailAdd{yijian.du@thep.lu.se~~~~~~~~~~~~~~~~~~~~~~~~~~}
\emailAdd{~~~~~~~physics@hqliu.com}
\emailAdd{shuyili19921123@hotmail.com}

\date{\today}
\abstract{In this note, we investigate relations between tree-level off-shell currents in nonlinear sigma model. Under Cayley parametrization, all odd-point currents vanish. We propose and prove a generalized $U(1)$ identity for even-point currents. The off-shell $U(1)$ identity given in \cite{Chen:2013fya} is a special case of the generalized identity studied in this note. The on-shell limit of this identity is equivalent with the on-shell KK relation. Thus this relation provides the full off-shell correspondence of tree-level KK relation in nonlinear sigma model.
}


\begin{document}
%%%%%%%%%%%%%%%%%%%%%%%%%%%%%%%%%%%%%%%%%%%%%%%%%%%%%%%%%%%%%%%%%%
\maketitle

\section{Introduction}

Color-kinematic duality (or BCJ duality) \cite{Bern:2008qj} was discovered by Bern, Carrasco and Johansson in 2008.
This duality states that Yang-Mills amplitudes can be written in a so-called BCJ formula where kinematic factors share the same algebraic properties (including antisymmetry and Jacobi identity) with color factors. BCJ duality implies relations between color-ordered amplitudes at tree-level. Specifically,
 the antisymmetry implies Kleiss-Kuijf relation (KK relation) \cite{Kleiss:1988ne}, while the Jacobi-identity implies Bern-Carrasco-Johansson (BCJ) relation \cite{Bern:2008qj}.
 With these relations, one can reduce the number of independent tree-level color-ordered amplitudes to $(n-3)!$.
Both KK and BCJ relations have been proven in string theory \cite{BjerrumBohr:2009rd,Stieberger:2009hq} and field theory \cite{DelDuca:1999rs,Feng:2010my,Tye:2010kg,Chen:2011jxa,Cachazo:2012uq}.

To understand the duality, further efforts including the loop-level BCJ duality \cite{Bern:2010yg, BjerrumBohr:2011xe, Carrasco:2011mn, Boels:2011tp,Boels:2011mn, Bern:2012uf, Carrasco:2012ca, Du:2012mt, Oxburgh:2012zr, Saotome:2012vy, Boels:2012ew, Carrasco:2012ca, Boels:2013bi, Bjerrum-Bohr:2013iza, Bern:2013yya,  Nohle:2013bfa}, the construction of BCJ numerators (by pure spinor string method \cite{Mafra:2011kj}, by kinematic algebra \cite{Monteiro:2011pc, BjerrumBohr:2012mg, Fu:2012uy, Monteiro:2013rya}, with relabeling symmetry \cite{Broedel:2011pd,Fu:2014pya,Naculich:2014rta} and from scattering equations \cite{Cachazo:2013gna, Cachazo:2013hca, Cachazo:2013iea,Naculich:2014rta,Naculich:2014naa}) as well as the dual trace-factors \cite{Bern:2011ia,Du:2013sha,Fu:2013qna,Du:2014uua,Naculich:2014rta} have been made. In another direction, one may wonder whether the BCJ duality and the amplitude relations implied by the duality exist in other theories. An interesting example is the duality and relations in three dimensional supersymmetric theories with $3$-algebra \cite{Bargheer:2012gv}. Another interesting extension is the amplitude relations in nonlinear sigma model with traditional $U(N)$ Lie algebra \cite{Chen:2013fya}.

In \cite{Chen:2013fya}, the authors  proved the $U(1)$ identity and the fundamental BCJ relation for three-level currents with  one off-shell leg. The on-shell versions of these two relations were obtained by taking on-shell limit of the off-shell leg.
Using the method for generating general on-shell KK and BCJ relations by the fundamental ones \cite{Ma:2011um}, one can obtain all the general on-shell KK and BCJ relations which have the same formulae with the corresponding relations in Yang-Mills theory.

Although all the on-shell versions of KK and BCJ relations for tree-level amplitudes in nonlinear sigma model have been proven in \cite{Chen:2013fya}, only two special off-shell relations, namely $U(1)$ identity and fundamental BCJ relation, have been studied. These off-shell relations  do not share the same formulae with those in Yang-Mills theory \cite{Chen:2013fya}. Actually, in Yang-Mills theory, there have been suggested (all leg) off-shell KK relations \cite{Du:2011js} which have the same formulae with the corresponding on-shell relations. No BCJ relation for off-shell currents in Yang-Mills theory was found\footnote{Only the off-shell BCJ relation for $\phi^3$ colored scalar theory was proposed \cite{Du:2011js}}.

A question is whether we can find the full off-shell extensions of the general on-shell KK and BCJ relations in nonlinear sigma model. There are several possible ways to think about this question. One way is to construct the BCJ formula in nonlinear sigma model and apply the algebraic properties to the kinematic factors. The main obstacles for this approach are the infinite number of vertices and the existence of off-shell leg. Another attempt is to generate all off-shell relations from the known off-shell $U(1)$ identity and  off-shell fundamental BCJ relation. However, the existence of the off-shell leg again becomes the main trouble.

In this note, we propose a generalized $U(1)$ identity for even-point off-shell currents $J(\sigma)$ in nonlinear sigma model. As already shown in the papers \cite{Kampf:2012fn, Kampf:2013vha}, under Cayley parametrization, the odd-point currents (with even numbers of on-shell legs and one off-shell leg) have to vanish \cite{Kampf:2012fn, Kampf:2013vha}. The generalized $U(1)$ identity for even-point currents (with odd numbers of on-shell legs and one off-shell leg) is given by
\bea
&&\Sl_{\sigma\in OP(\{\alpha_1,\dots,\alpha_r\}\bigcup\{\beta_1,\dots,\beta_{s}\})}J(\sigma)\nn
&=&\Sl_{D\in\text{Divisions of $\{\alpha\}$, $\{\beta\}$} \atop \text{s.t.}, |R_D-S_D|=1
}\left({1\over 2F^2}\right)^{R_{D}+S_D-1\over 2}J(A_{1})\dots J(A_{R_D})J(B_{1})\dots J(B_{S_D}).\Label{off-shell-gen-U(1)}
\eea
On the left hand side of \eqref{off-shell-gen-U(1)}, we summed over all the ordered permutations $OP(\{\alpha\}\bigcup\{\beta\})$ with keeping the relative orders in each set. For example, in $OP(\{\alpha_1,\alpha_2\}\bigcup\{\beta_1,\beta_2\})$, we have permutations $(\alpha_1,\alpha_2,\beta_1,\beta_2)$, $(\alpha_1,\beta_1,\alpha_2,\beta_2)$, $(\alpha_1,\beta_1,\beta_2,\alpha_2)$, $(\beta_1,\alpha_1,\alpha_2,\beta_2)$, $(\beta_1,\alpha_1,\beta_2,\alpha_2)$, $(\beta_1,\beta_2,\alpha_1,\alpha_2)$. On the right hand side, we have summed over all the possible divisions $D$ of $\{\alpha\}$ and $\{\beta\}$ into ordered subsets $\{A_1\},\dots,\{A_{R_D}\}$ and $\{B_1\},\dots,\{B_{S_D}\}$ with odd numbers of elements in each subset. The numbers of subsets $R_D$ and $S_D$  for given division $D$ should satisfy $|R_D-S_D|=1$. For example, if we have three elements in the $\{\alpha\}$ set and four elements in the $\{\beta\}$ set, we have
\begin{itemize}
\item two $(1,2)$ divisions with $\{\alpha\}\to\{\alpha_1,\alpha_2,\alpha_3\}$ and $\{\beta\}\to\{\beta_1\},\{\beta_2,\beta_3,\beta_4\}$ or $\{\beta\}\to\{\beta_1,\beta_2,\beta_3\},\{\beta_4\}$
    \item two $(3,2)$ divisions with $\{\alpha\}\to\{\alpha_1\},\{\alpha_2\},\{\alpha_3\}$ and $\{\beta\}\to\{\beta_1\},\{\beta_2,\beta_3,\beta_4\}$ or $\{\beta\}\to\{\beta_1,\beta_2,\beta_3\},\{\beta_4\}$
        \item one $(3,4)$ division  with $\{\alpha\}\to\{\alpha_1\},\{\alpha_2\},\{\alpha_3\}$ and $\{\beta\}\to\{\beta_1\},\{\beta_2\},\{\beta_3\},\{\beta_4\}$.
\end{itemize}
A special case of the off-shell generalized $U(1)$ identity \eqref{off-shell-gen-U(1)} is $r=1$ (or $s=1$). In this case, $R$, $S$ (or $S$, $R$) have to be $1$, $2$ respectively and we arrive at the $U(1)$ identity proven in \cite{Chen:2013fya} (see \eqref{off-shell-U(1)}). When multiplying a $p_1^2\to 0$ to the right hand side of the relation \eqref{off-shell-gen-U(1)}, we just arrive at the corresponding on-shell relation for color-ordered amplitudes $A$\footnote{The on-shell generalized U(1) identity in Yang-Mills theory was firstly proposed in \cite{Berends:1988zn} }
\bea
&&\Sl_{\sigma\in OP(\{\alpha_1,\dots,\alpha_r\}\bigcup\{\beta_1,\dots,\beta_{s}\})}A(1,\{\sigma\})=0,
\eea
which has been shown to be equivalent with the on-shell KK relation  \cite{Du:2011js}.

To prove the off-shell identity \eqref{off-shell-gen-U(1)}, we first study the eight-point identity with $r=3, s=4$ by explicit calculations with Berends-Giele recursion. Because of the complexity, it seems impossible to extend the  calculation directly to a general proof. Instead, we redefine the coefficients for products of subcurrents level by level. After this redefinition, all the divisions $D$ with $R_D+S_D<r+s$ have the right coefficients in \eqref{off-shell-gen-U(1)}. Then we only need to prove that the coefficient for $(r,s)$ division has the right form.
By combining the $U(1)$ identity with a generalized $U(1)$ identity with fewer $\alpha$'s, we obtain a set of equations which are finally used to determine the $(r,s)$ coefficient.

The structure of this note is following. In section 2, we review the Feynman rules, Berends-Giele recursion and the $U(1)$ identity proved in the paper \cite{Chen:2013fya}.
In section 3, we study the generalized $U(1)$ identity with three elements in $\{\alpha\}$ and four elements in $\{\beta\}$  by  Berends-Giele recursion directly. It will be quite hard to extend this calculation to a general proof. In section 4, we provide another approach by redefining the coefficients of divisions with $R_D+S_D<r+s$ step by step. After these redefinitions, all  divisions with $R_D+S_D<r+s$ already have the right coefficients. We then prove that the coefficient for $(r,s)$ division also has the right form. At last, we conclude this work in section 5.

\section{Preparation: Feynman rules, Berends-Giele recursion and $U(1)$ identity}
In this section, we review Feynman rules, the Berends-Giele recursion and the $U(1)$ identity in nonlinear sigma model\footnote{Parts of this section overlap with the section 2 of \cite{Chen:2013fya}}. Most of the notations follow the recent papers \cite{Kampf:2012fn,Kampf:2013vha}.

%%%%%%%%%%%%%%%%%%%%%
\subsection{Feynman rules}
%%%%%%%%%%%%%%%%%%%%%

{~~~~~\emph {Lagrangian}}

The Lagrangian of $U(N)$ non-linear sigma model is
\bea
\mathcal{L}={F^2\over 4}\Tr (\partial_{\mu}U\partial^{\mu}U^{\dagger}),
\eea
where $F$ is a constant. Using Caylay parametrization as in \cite{Kampf:2012fn,Kampf:2013vha}, the $U$ is defined by
\bea
U=1+2\Sl_{n=1}^{\infty}\left({1\over 2F}\phi\right)^n,~~~~\label{Cayley}
\eea
where $\phi=\sqrt{2}\phi^at^a$ and $t^a$ are generators of $U(N)$ Lie algebra.

{\emph {Trace form of color decomposition}}

The full tree amplitudes can be given by trace form decomposition
\bea
M(1^{a_1},\dots,n^{a_n})=\Sl_{\sigma\in S_{n-1}}\Tr(T^{a_{1}}T^{a_{\sigma_2}}\dots T^{a_{\sigma_n}})A(1,\sigma).\label{Trace form}
\eea
Since traces have cyclic symmetry, the color-ordered amplitudes $A$ also satisfy cyclic symmetry
\bea
A(1,2,\dots,n)=A(n,1,\dots,n-1).\label{Cyclic symmetry}
\eea

\emph{Feynman rules for color-ordered amplitudes}

Vertices in color-ordered Feynman rules under Cayley parametrization \eqref{Cayley} are
\bea
V_{2n+1}=0, V_{2n+2}=\left(-{1\over 2F^2}\right)^n\left(\Sl_{i=0}^np_{2i+1}\right)^2=\left(-{1\over 2F^2}\right)^n\left(\Sl_{i=0}^np_{2i+2}\right)^2.\Label{Feyn-rules}
\eea
Here, $p_j$ denotes the momentum of the leg $j$; momentum conservation has been considered.

%%%%%%%%%%%%%%%%%%%%%%%%
\subsection{Berends-Giele recursion}
%%%%%%%%%%%%%%%%%%%%%%%%
In the Feynman rules given above, one can construct tree-level currents\footnote{In this paper, an $n$-point current is mentioned as the current with $n-1$ on-shell legs and one off-shell leg.} through Berends-Giele recursion
\bea
&&J(2,...,n)\nn
&=&\frac{i}{P_{2,n}^2}\Sl_{m=4}^n\Sl_{1=j_0<j_1<\cdots<j_{m-1}=n}i V_{m}(p_1=-P_{2,n},P_{j_0+1,j_1},\cdots,P_{j_{m-2}+1,n})\times\prod\limits_{k=0}^{m-2} J(j_k+1,\cdots,j_{k+1}),\Label{B-G}\nn
\eea
where $p_1=-P_{2,n}=-(p_2+p_3+\dots+p_n)$ is the momentum of the off-shell leg $1$. The starting point of this recursion is $J(2)=J(3)=\dots=J(n)=1$.

 Since there is at least one odd-point vertex for current with odd-point lines (including the off-shell line) and the odd-point vertices are zero, we always have
\bea
J(2,\dots,2m+1)=0,
\eea
for $(2m+1)$-point amplitudes.
The currents with even points in general are nonzero and are built up by only odd numbers of even-point sub-currents. Since odd-point currents have to vanish,
in all following sections of this paper, we just need to discuss on the relations among even-point currents.

%%%%%%%%%%%%%%%%%%%%%%%%%%%%%
\subsection{The off-shell versions of $U(1)$ identity}
%%%%%%%%%%%%%%%%%%%%%%%%%%%%
In \cite{Chen:2013fya}, the authors have proven the $U(1)$ identity for off-shell currents in nonlinear sigma model.
The identity is
\bea
\Sl_{\sigma\in OP(\{\alpha_1\}\bigcup\{\beta_1,\dots,\beta_{2m}\})}J(\sigma)={1\over 2F^2}\Sl_{divisions\{\beta\}\rightarrow\{B_{1}\},\{B_{2}\}}J(B_{1})J(B_{2}),\Label{off-shell-U(1)}
\eea
where on the left hand side, we summed over the permutations in $\{\alpha_1\}\bigcup\{\beta_1,\dots,\beta_{2m}\}$ with keeping the relative order in the $\beta$ set. On the right hand side, we summed over the divisions of $\{\beta\}$ into two ordered subsets.

\section{Direct calculation of an eight-point example}

We have checked the generalized $U(1)$ identity \eqref{off-shell-gen-U(1)} for four- and six-point currents. In the four-point case,
we only have $r=1,s=2$ and  $r=2,s=1$, which are $U(1)$ identities \eqref{off-shell-U(1)}. In the six-point case, $r=1,s=4$ and $r=4,s=1$ are also $U(1)$ identities \eqref{off-shell-U(1)}. The new relations for six-point currents are the cases with $r=2, s=3$ and $r=3,s=2$, where the later case can be obtained from the former one by exchanging the roles of $\alpha$ and $\beta$. We just skip all the calculations of four- and six-point identities and show a more complicated eight-point example.

\begin{table}
{\begin{tabular}{|c|c|c|c|}
  \hline
  Divisions & Type-1 & Type-2 & Type-3 \\ \hline
  $\{\alpha_1\}\{\alpha_2\}\{\alpha_3\}\{\beta_1\}\{\beta_2\}\{\beta_3\}\{\beta_4\}$ & $s_{\alpha_1\alpha_3}+s_{\beta_1\beta_3}+s_{\beta_2\beta_4}$ & $0$ & $p_1^2$ \\\hline
  $\{\alpha_1\}\{\alpha_2\}\{\alpha_3\}\{\beta_1,\beta_2,\beta_3\}\{\beta_4\}$ & $-s_{\alpha_1\alpha_3}$ & $-\left(p_{\beta_1}+p_{\beta_2}+p_{\beta_3}\right)^2$ & $p_1^2$ \\\hline
  $\{\alpha_1\}\{\alpha_2\}\{\alpha_3\}\{\beta_1\}\{\beta_2,\beta_3,\beta_4\}$ & $-s_{\alpha_1\alpha_3}$ & $-\left(p_{\beta_2}+p_{\beta_3}+p_{\beta_4}\right)^2$ & $p_1^2$ \\\hline
 $\{\alpha_1,\alpha_2,\alpha_3\}\{\beta_1\}\{\beta_2\}\{\beta_3\}\{\beta_4\}$ & $-s_{\beta_1\beta_3}-s_{\beta_2\beta_4}$ & $-\left(p_{\alpha_1}+p_{\alpha_2}+p_{\alpha_3}\right)^2$ & 0 \\\hline
 $\{\alpha_1,\alpha_2,\alpha_3\}\{\beta_1,\beta_2,\beta_3\}\{\beta_4\}$ & 0 & $\left(p_{\alpha_1}+p_{\alpha_2}+p_{\alpha_3}\right)^2+\left(p_{\beta_1}+p_{\beta_2}+p_{\beta_3}\right)^2$ & $p_1^2$ \\\hline
   $\{\alpha_1,\alpha_2,\alpha_3\}\{\beta_1\}\{\beta_2,\beta_3,\beta_4\}$ & 0 & $\left(p_{\alpha_1}+p_{\alpha_2}+p_{\alpha_3}\right)^2+\left(p_{\beta_2}+p_{\beta_3}+p_{\beta_4}\right)^2$ & $p_1^2$ \\\hline
\end{tabular}}
\caption{The coefficients of eight-point case in general can be classified into three types. Here we omit the coupling constants for convenience.}\label{8pt-gen-U(1)-cancelation}
\end{table}

We take the eight-point identity with three elements in $\{\alpha\}$ and four elements in $\{\beta\}$ as an example. The explicit form of the identity \eqref{off-shell-gen-U(1)} with $r=3,s=4$ is
\bea
&&\Sl_{\sigma\in OP(\{\alpha_1,\alpha_2,\alpha_3\}\bigcup\{\beta_1,\beta_2,\beta_3,\beta_4\})}J(\sigma)\nn
&=&{1\over 2F^2}\left[J(\alpha_1,\alpha_2,\alpha_3)J(\beta_1)J(\beta_2,\beta_3,\beta_4)+J(\alpha_1,\alpha_2,\alpha_3)J(\beta_1,\beta_2,\beta_3)J(\beta_4)\right]\nn
&&+\left({1\over 2F^2}\right)^2\left[J(\alpha_1)J(\alpha_2)J(\alpha_3)J(\beta_1)J(\beta_2,\beta_3,\beta_4)+J(\alpha_1)J(\alpha_2)J(\alpha_3)J(\beta_1,\beta_2,\beta_3)J(\beta_4)\right]\nn
&&+\left({1\over 2F^2}\right)^3J(\alpha_1)J(\alpha_2)J(\alpha_3)J(\beta_1)J(\beta_2)J(\beta_3)J(\beta_4).\Label{8pt-gen-U(1)-example}
\eea
\begin{figure}
  \centering
 \includegraphics[width=0.65\textwidth]{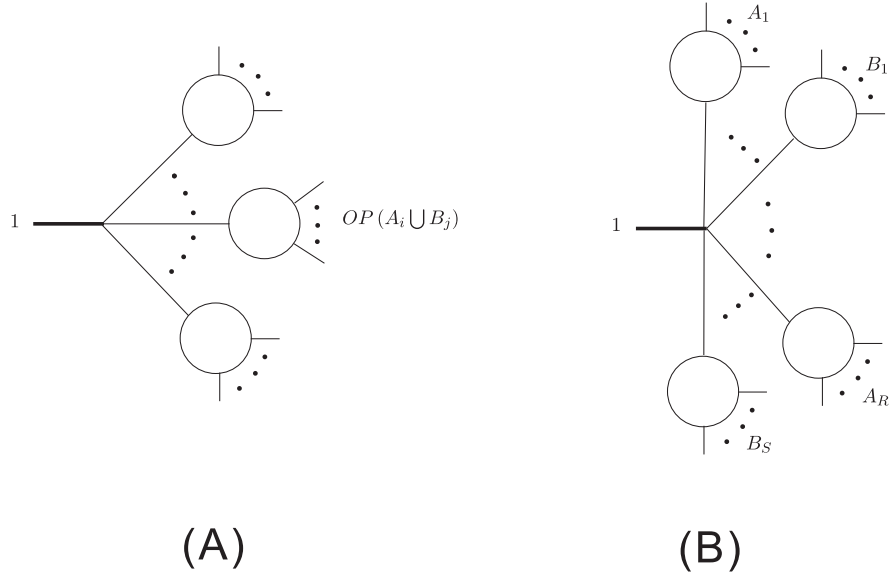}
 \caption{Two classes of diagrams: (A) diagrams containing substructures of generalized $U(1)$ identity such as $\sigma\in OP(A_i\bigcup B_j)$ where $A_i$ and $B_j$ denote ordered subsets of $\{\alpha\}$ and $\{\beta\}$. (B) diagrams with each subcurrent containing only $\{\alpha\}$ elements or $\{\beta\}$ elements.}\label{Two classes of diagrams.}
\end{figure}
%%%

To prove this identity, we use Berends-Giele recursion \eqref{B-G} to express all the currents on the left hand side of \eqref{8pt-gen-U(1)-example} by six- and four-point subcurrents.
Then we collect the terms with a same vertex connected to the off-shell leg $1$. After summing all the possible diagrams in each collection, the left hand side of \eqref{8pt-gen-U(1)-example} is expressed by
\begin{itemize}
\item diagrams containing six-point and (or) four-point substructures of generalized $U(1)$-identity (see Fig. \ref{Two classes of diagrams.}(A))

\item  diagrams with neither six-point nor four-point substructure (see Fig. \ref{Two classes of diagrams.}(B)).
 \end{itemize}
Remembering that the identity \eqref{off-shell-gen-U(1)} is satisfied by four- and six-point currents, we apply these lower-point identities to the four- and six-point substructures in the first class of diagram. Then diagrams in the first class are rewritten in terms of products of subcurrents containing only $\alpha$ or $\beta$ elements. Since the second class of diagram does not have any substructure, it is already expressed by products of subcurrents containing only $\alpha$ or $\beta$ elements.
 After this reduction, for a given product of subcurrents (or in other words, given division of $\{\alpha\}$ set and $\{\beta\}$ set), we collect the coefficients together. Thus the left hand side of \eqref{8pt-gen-U(1)-example} is written as
\bea
&&\left(1\over 2F^2\right)^3\left[\left(s_{\alpha_1\alpha_3}+s_{\beta_1\beta_3}+s_{\beta_2\beta_4}\right)+p_1^2\right]{1\over p_1^2}J(\alpha_1)J(\alpha_2)J(\alpha_3)J(\beta_1)J(\beta_2)J(\beta_3)J(\beta_4)\nn
&+&\left(1\over 2F^2\right)^2\left[-s_{\alpha_1\alpha_3}-\left(p_{\beta_1}+p_{\beta_2}+p_{\beta_3}\right)^2+p_1^2\right]{1\over p_1^2}J(\alpha_1)J(\alpha_2)J(\alpha_3)J(\beta_1,\beta_2,\beta_3)J(\beta_4)\nn
&+&\left(1\over 2F^2\right)^2\left[-s_{\alpha_1\alpha_3}-\left(p_{\beta_2}+p_{\beta_3}+p_{\beta_4}\right)^2+p_1^2\right]{1\over p_1^2}J(\alpha_1)J(\alpha_2)J(\alpha_3)J(\beta_1)J(\beta_2,\beta_3,\beta_4)\nn
&+&\left(1\over 2F^2\right)^2\left[\left(-s_{\beta_1\beta_3}-s_{\beta_2\beta_4}\right)-\left(p_{\alpha_1}+p_{\alpha_2}+p_{\alpha_3}\right)^2\right]{1\over p_1^2}J(\alpha_1,\alpha_2,\alpha_3)J(\beta_1)J(\beta_2)J(\beta_3)J(\beta_4)\nn
&+&\left(1\over 2F^2\right)\left[\left(p_{\alpha_1}+p_{\alpha_2}+p_{\alpha_3}\right)^2+\left(p_{\beta_1}+p_{\beta_2}+p_{\beta_3}\right)^2+p_{1}^2\right]{1\over p_1^2}J(\alpha_1,\alpha_2,\alpha_3)J(\beta_1,\beta_2,\beta_3)J(\beta_4)\nn
&+&\left(1\over 2F^2\right)\left[\left(p_{\alpha_1}+p_{\alpha_2}+p_{\alpha_3}\right)^2+\left(p_{\beta_2}+p_{\beta_3}+p_{\beta_4}\right)^2+p_{1}^2\right]{1\over p_1^2}J(\alpha_1,\alpha_2,\alpha_3)J(\beta_1)J(\beta_2,\beta_3,\beta_4),
\eea
where $s_{ij}\equiv(p_i+p_j)^2$.
Coefficients for each division can be classified into three types (see \ref{8pt-gen-U(1)-cancelation}).
A type-2 coefficient always cancels with a propagator of a subcurrent and divides the subcurrent into new subcurrents.
For example, the coefficient in type-2 term on the second line is $-(p_{\beta_1}+p_{\beta_2}+p_{\beta_3})^2$ which reduce the current $J(\beta_1,\beta_2,\beta_3)$ to $-\left(1\over 2F^2\right)s_{\beta_1\beta_3}J(\beta_1)J(\beta_2)J(\beta_3)$. Thus this part of contribution cancels with the second term of the type-1 coefficient of $(3,4)$ division. Similarly, other type-2 terms also cancel with type-1 terms for divisions with larger $R_D+S_D$. All the type-1 and type-2 terms cancel out in this way. Only the type-3 terms are left and give the right hand side of the eight-point identity \eqref{8pt-gen-U(1)-example}.

\section{Proof of the generalized $U(1)$ identity for off-shell currents}
%%%
\begin{figure}
  \centering
 \includegraphics[width=0.35\textwidth]{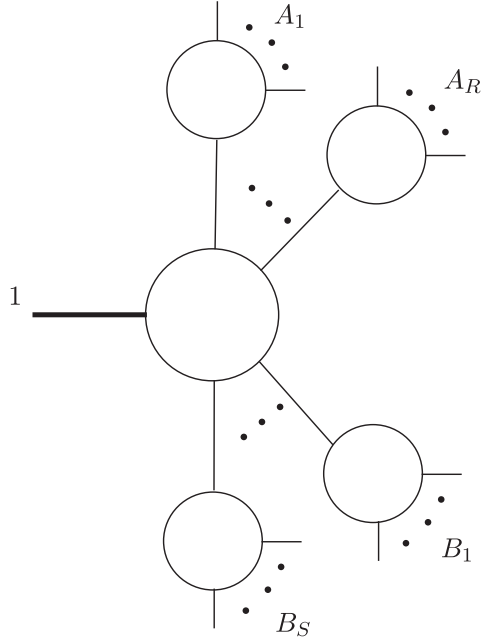}
 \caption{We redefine the coefficients for $R+S<r+s$ divisions such that they are the right ones as in the general form \eqref{off-shell-gen-U(1)}. Then we solve the $R=r$, $S=s$ coefficient.}\label{CurrentDivsion}
\end{figure}
%%%

In the previous section, we have provided a  direct approach to an eight-point example by Berends-Giele recursion. Although the coefficients in the example were shown to have a good pattern (see table \ref{8pt-gen-U(1)-cancelation}), it will be quite hard to extend the calculation to a general proof. One reason is that we will encounter many different lower-point substructures of the identity \eqref{off-shell-gen-U(1)} when the number of $\{\alpha\}$ elements grows.  Thus we have to prove the general formula \eqref{off-shell-gen-U(1)} in a different way. In this section, we will show a general proof of the identity \eqref{off-shell-gen-U(1)}. The main idea is following:
\begin{itemize}
\item As we have done in the eight-point example, we write the left hand side of the identity \eqref{off-shell-gen-U(1)} by Berends-Giele recursion and collect the diagrams with a same vertex attached to the off-shell leg $1$ (See Fig.  \ref{Two classes of diagrams.}(A) and (B)). Reducing the substructures  by lower-point identities and putting the coefficients corresponding to each product of subcurrents together, we express the left hand side of \eqref{off-shell-gen-U(1)} as follows
    \bea
   &&\Sl_{\sigma\in OP(\{\alpha_1,\dots,\alpha_r\}\bigcup\{\beta_1,\dots,\beta_{s}\})}J(\sigma)\nn
   &=&\Sl_{D\in Divisions}{1\over p_1^2}\left[\Sl_{i_{4,D}}V_4^{i_{4,D}}-\Sl_{i_{6,D}}V_6^{i_{6,D}}+\Sl_{i_{8,D}}V_8^{i_{8,D}}-\dots+(-1)^{R_D+S_D-1\over 2}\Sl_{i_{R_D+S_D+1,D}}V^{i_{R_D+S_D+1,D}}_{R_D+S_D+1}\right]\nn
   &&~~~~~~~~~~~~~~~~~~~~~~~~~~~~~~~~~~~~~~~~~~~~~~~~~~~~~~~~~~~~~~~~~~~~~~\times J(A_1)\dots J(A_{R_D})J(B_1)\dots J(B_{S_D}),\Label{off-shell-gen-U(1)-0}
    \eea
   where $V_l^{i_{l,D}}$ denote the $l$-point vertices which contribute to the division $D$ and $\Sl_{i_{4,D}}$ means that we sum over all such $l$-point vertices.
   The prefactor  $(-1)^{{l-1\over 2}}$ of $l$-point vertex comes from the factor $\left(-{1\over 2F^2}\right)^n$ in the Feynman rules \eqref{Feyn-rules}.

\item We show that the expression obtained in the above step can be rearranged (figure \ref{CurrentDivsion}) into the following formula
      \bea
      &&\Sl_{\sigma\in OP(\{\alpha_1,\dots,\alpha_r\}\bigcup\{\beta_1,\dots,\beta_{s}\})}J(\sigma)\nn
      &=&\Sl_{D\in \text{Divisions of}~\{\alpha\},\{\beta\}\atop R_D+S_D<r+s}
\left({1\over 2F^2}\right)^{R_D+S_D-1\over 2}\delta(|R_D-S_D|-1)J(A_{1})\dots J(A_{R_D})J(B_{1})\dots J(B_{S_D})\nn
&&~~~~~~~~~~~~~~~~~~~~~~~~~~~~~~~~~~~~~+\left({1\over 2F^2}\right)^{r+s-1\over 2}{\cal V}^{(r,s)}J(\alpha_{1})\dots J(\alpha_r)J(\beta_{1})\dots J(\beta_s),\Label{off-shell-gen-U(1)-1}
      \eea
      where ${\cal V}^{(r,s)}$ is the dimensionless coefficient for the $(r,s)$ division.
      The first term of \eqref{off-shell-gen-U(1)-1} is given by sum of divisions $D$ $(R_D+S_D<r+s)$ which already have the correct coefficients in \eqref{off-shell-gen-U(1)}. Thus we only need to prove that the coefficient  ${\cal V}^{(r,s)}$ in the second term of  \eqref{off-shell-gen-U(1)-1} also has the right expression in \eqref{off-shell-gen-U(1)}, i.e., ${\cal V}^{(r,s)}=\delta(|r-s|-1)$.
      %Thus we only need to prove ${\cal V}^{(r,s)}=\delta(|r-s|-1)$ which gives rise to the right coefficient for the $(r,s)$ division in \eqref{off-shell-gen-U(1)}.
      %We leave the proof of ${\cal V}^{(r,s)}=\delta(|r-s|-1)$ to the next step and assume that for lower-point cases with $r'$ $\alpha$'s and $s'$ $\beta$'s $(r'+s'<r+s)$, we already have the right coefficients ${\cal V}^{(r',s')}$. Then we can use the lower-point ${\cal V}^{(r',s')}$ to prove the validity of
      %\eqref{off-shell-gen-U(1)-1}.
      \item  The undetermined coefficient ${\cal V}^{(r,s)}$  has the general form ${1\over p_1^2}\left(\Sl_{i,j}c_{ij}s_{ij}\right)$ with appropriate $c_{ij}$. By combining an $U(1)$ identity and a generalized $U(1)$ identity with fewer $\alpha$'s, we can prove that ${\cal V}^{(r,s)}=\delta(|r-s|-1)$.
      Therefore, the generalized $U(1)$ identity \eqref{off-shell-gen-U(1)} for off-shell currents is proved.
\end{itemize}
In the remainder of this section, we will show the left hand side of \eqref{off-shell-gen-U(1)} can be rearranged into \eqref{off-shell-gen-U(1)-1} and then solve ${\cal V}^{(r,s)}$.

\subsection{Proof of the validity of  \eqref{off-shell-gen-U(1)-1} with an undetermined coefficient ${\cal V}^{(r,s)}$}

Now we show that the left hand side of the off-shell generalized $U(1)$ identity \eqref{off-shell-gen-U(1)} can be rearranged into the form \eqref{off-shell-gen-U(1)-1}. We start from several examples.
%%%
\begin{figure}
  \centering
 \includegraphics[width=0.8\textwidth]{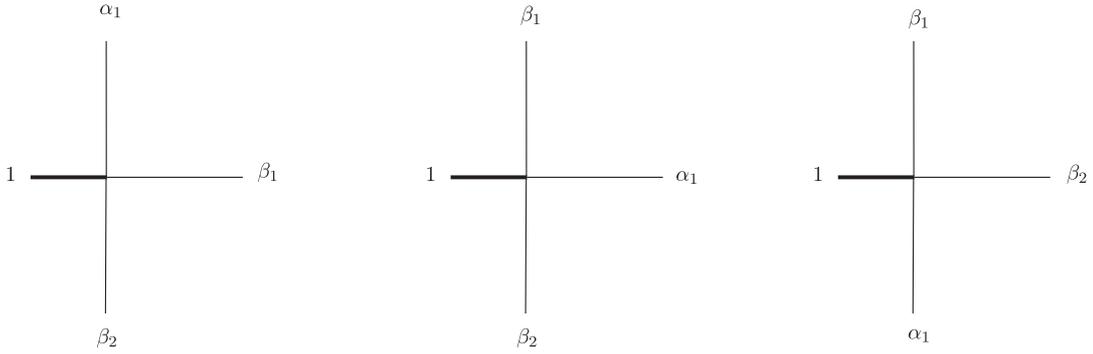}
 \caption{Diamgrams contributing to four-point identity.}\label{4ptDiagrams}
\end{figure}
%%%

{\bf Four-point example:}  The four-point example is the four-point $U(1)$ identity (see \cite{Chen:2013fya}). By explicit calculation, this is given by the sum of three diagrams in Fig. \ref{4ptDiagrams}, i.e.,
\bea
\Sl_{\sigma\in OP(\{\alpha_1\}\bigcup\{\beta_1,\beta_2\})}J(\sigma)=\left(1\over 2F^2\right)J(\alpha_1)J(\beta_1)J(\beta_2).\Label{4pt-U(1)}
\eea
The identity with two $\alpha$'s and one $\beta$ can be obtained by exchanging the roles of $\alpha$ and $\beta$.
%%%
\begin{figure}
  \centering
 \includegraphics[width=0.8\textwidth]{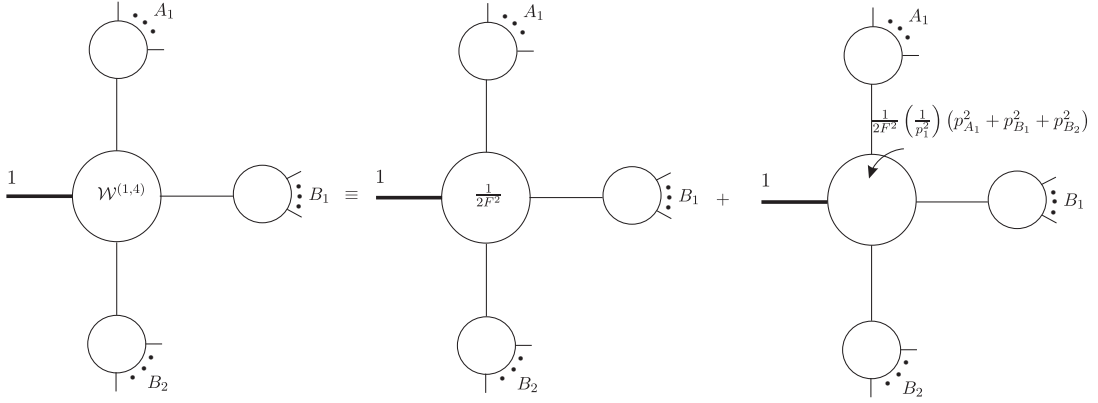}
 \caption{The off-shell extension of four-point identity. Here we absorb the ${1\over p_1^2}$ corresponding to the off-shell leg $1$ into the coefficients for convenience. }\label{4pt-off-shell-extension}
\end{figure}
%%%

Before giving the next example, let us have a look at an off-shell extension of the right hand side of \eqref{4pt-U(1)}, i.e., we replace the three on-shell legs $\alpha_1$, $\beta_1$ and $\beta_2$ in Fig. \ref{4ptDiagrams} by three off-shell currents $J(A_1)$, $J(B_1)$ and $J(B_2)$ correspondingly. From Feynman rules \eqref{Feyn-rules}, the coefficient of $J(A_1)J(B_1)J(B_2)$ is written as a linear combination of ${1\over p_1^2}(p_{i}\cdot p_j)$, where $i$, $j$ can be either one of $A_1$, $B_1$, $B_2$; we use $p_{A_i}$, $p_{B_i}$ to denote the sum of momenta of elements in  $A_i$, $B_i$ respectively. Then we consider the sum of diagrams with the off-shell leg $1$ connected to a four-point vertex whose other three legs are attached to the currents $J(A_1)$, $J(B_1)$ and $J(B_2)$.
In general, we should have
\bea
\Sl_{\sigma\in OP(\{A_1\}\bigcup\{B_1,B_2\})}J^{(4)}(\sigma)={\cal W}^{(1,2)}J(A_1)J(B_1)J(B_2),\Label{4pt-U(1)-off-shell}
\eea
with
\bea
{\cal W}^{(1,2)}\equiv\left(1\over 2F^2\right)\left[1+{1\over p_1^2}\left(a^{(1,2)}_1p_{A_1}^2+b^{(1,2)}_1p_{B_1}^2+b^{(1,2)}_2p_{B_2}^2\right)\right].\Label{W(1,2)}
\eea
Here, $J^{(4)}(\sigma)$ denote the diagrams with the four-point vertices connected to $1$, $J(A_1)$, $J(B_1)$ and $J(B_2)$ and $a^{(1,2)}_1$, $b^{(1,2)}_1$ and $b^{(1,2)}_2$ are some constant coefficients. The equation \eqref{4pt-U(1)-off-shell} is the only possible formula of all-leg-off-shell extension of the four-point identity \eqref{4pt-U(1)} for one-leg-off-shell currents. This is because when replacing the currents $J(A_1)$, $J(B_1)$ and $J(B_2)$ by on shell legs $\alpha_1$, $\beta_1$ and $\beta_2$, we have to return to \eqref{4pt-U(1)}. The coefficient thus can only be the sum of $\left(1\over 2F^2\right)$ and combinations of ${1\over p_1^2}p^2_{A_i}$, ${1\over p_1^2}p^2_{B_i}$, which vanish under on-shell limit. From the explicit calculation in \cite{Chen:2013fya}, we can see $a_1^{(1,2)}=b^{(1,2)}_1=b^{(1,2)}_2=1$. Thus the off-shell extension \eqref{W(1,2)} can be expressed by Fig. \ref{4pt-off-shell-extension}. We will encounter \eqref{4pt-U(1)-off-shell}, \eqref{W(1,2)} in higher-point cases.

{\bf Six-point example:} With the four-point identity in hand, let us consider the six-point example. The first six-point example is the $U(1)$-identity with only one $\alpha$, which has been understood. Now we consider the generalized identity with two $\alpha$'s
\bea
\Sl_{\sigma\in OP(\{\alpha_1,\alpha_2\}\bigcup\{\beta_1,\beta_2,\beta_3\})}J(\sigma)&=&{1\over 2F^2}J(\alpha_1)J(\alpha_2)J(\beta_1,\beta_2,\beta_3)+\left({1\over 2F^2}\right)^2J(\alpha_1)J(\alpha_2)J(\beta_1)J(\beta_2)J(\beta_3).\nn
\eea

{\bf Step-1} To prove this identity, we start from the left hand side. We use Berends-Giele recursion to express the currents on the left hand side. Then collect the diagrams together with same substructures of generalized $U(1)$-identity.
After reducing diagrams containing four-point substructures of $U(1)$-identity by \eqref{4pt-U(1)}, we collect the coefficients for given division of $\{\alpha\}$ and $\{\beta\}$. Then the left hand side of the six-point identity has the form
\bea
\Sl_{\sigma\in OP(\{\alpha_1,\alpha_2\}\bigcup\{\beta_1,\beta_2,\beta_3\})}J(\sigma)={\cal U}^{(2,3)}_1J(\alpha_1)J(\alpha_2)J(\beta_1)J(\beta_2)J(\beta_3)+{\cal U}^{(2,3)}_2J(\alpha_1)J(\alpha_2)J(\beta_1,\beta_2,\beta_3),\Label{6pt-1}
\eea
with ${\cal U}^{(2,3)}_1$ and ${\cal U}^{(2,3)}_2$ as coefficients. In general, ${\cal U}^{(2,3)}_1$ and ${\cal U}^{(2,3)}_2$  are written as sum of terms of the form ${1\over p_1^2}(p_{i}\cdot p_j)$, where ${1\over p_1^2}$ and $(p_{i}\cdot p_j)$ respectively come from the off-shell propagator and vertices (as shown in \eqref{off-shell-gen-U(1)-0}).
The second term in \eqref{6pt-1} is the $(2,1)$ division which can only get contribution from diagrams with the off-shell leg $1$ directly connected to four-point vertices whose other three lines are connected to $J(\alpha_1)$, $J(\alpha_2)$ and $J(\beta_1,\beta_2,\beta_3)$. The sum of such contributions is noting but the off-shell extension \eqref{4pt-U(1)-off-shell} in the four-point example with $A_1\to \{\beta_1,\beta_2,\beta_3\}$, $B_1\to \{\alpha_1\}$, $B_2\to \{\alpha_2\}$. Thus we have
\bea
{\cal U}^{(2,3)}_2=\left(1\over 2F^2\right)\left\{1+{1\over p_1^2}\left[a_1^{(1,2)}\left(p_{\beta_1}+p_{\beta_2}+p_{\beta_3}\right)^2\right]\right\},
\eea
where the on-shell conditions of $\alpha_1$ and $\alpha_2$ have been used.
%%%
\begin{figure}
  \centering
 \includegraphics[width=0.9\textwidth]{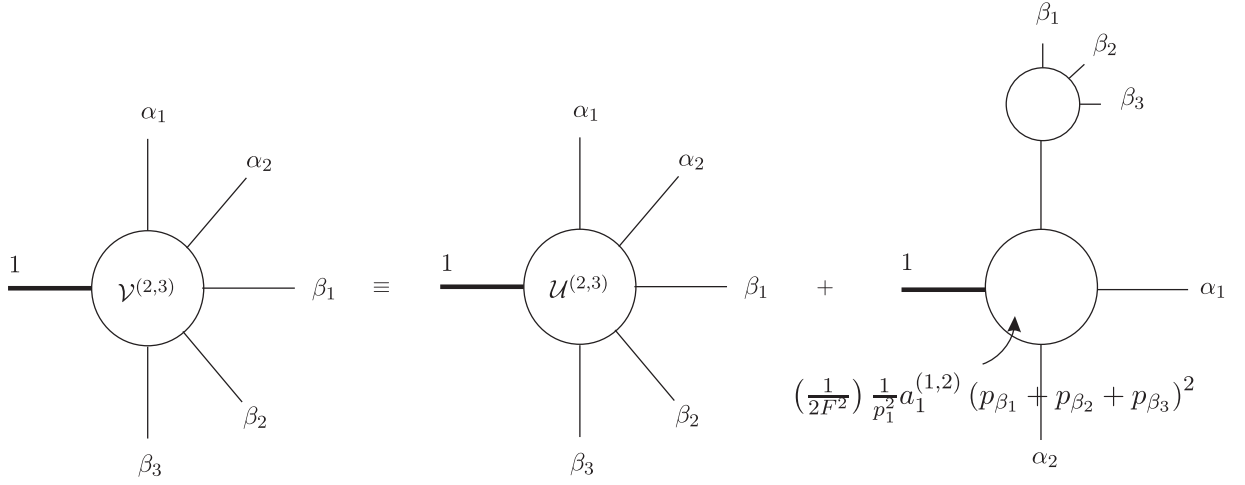}
 \caption{Redefinition of the coefficient of $(2,3)$ division for the identity with two $\alpha$'s and three $\beta$'s.}\label{redefining_coefficients_6pt}
\end{figure}
%%%
%%%
\begin{figure}
  \centering
 \includegraphics[width=0.9\textwidth]{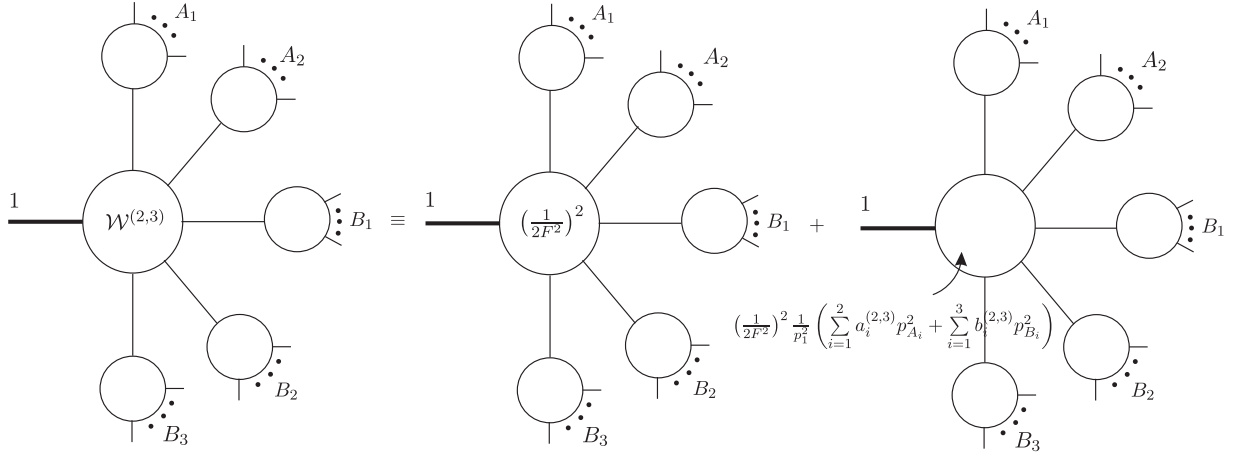}
 \caption{The off-shell extension of the $(2,3)$ division in the six-point identity with two $\alpha$'s and three $\beta$'s. }\label{6pt-off-shell-extension1}
\end{figure}
%%%
%%%
\begin{figure}
  \centering
 \includegraphics[width=0.8\textwidth]{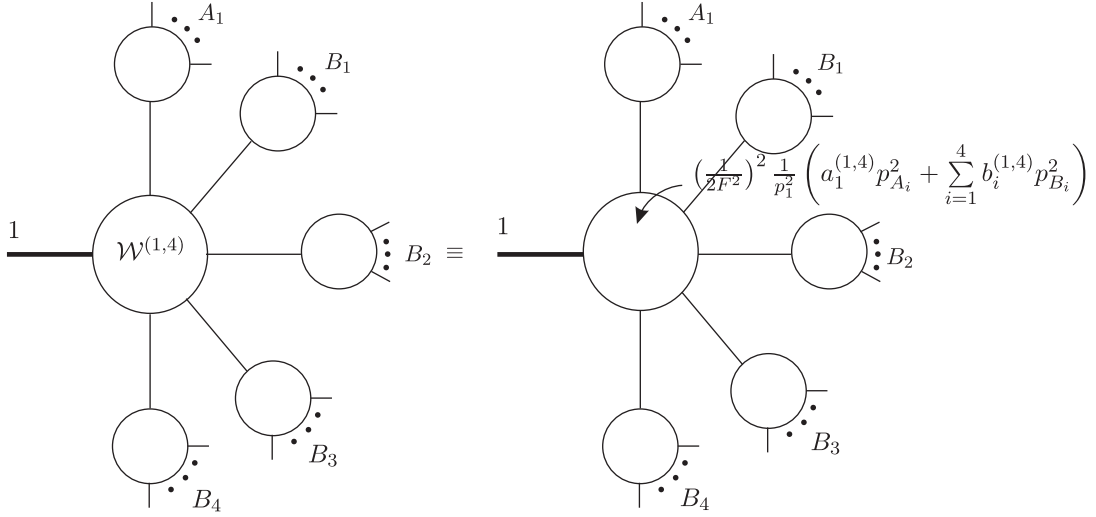}
 \caption{The off-shell extension of the $(1,4)$ division in the six-point identity with one $\alpha$ and four $\beta$'s. }\label{6pt-off-shell-extension2}
\end{figure}
%%%

{\bf Step-2} Since $\left(p_{\beta_1}+p_{\beta_2}+p_{\beta_3}\right)^2$ further reduces $J(\beta_1,\beta_2,\beta_3)$ to $J(\beta_1)J(\beta_2)J(\beta_3)$ with a coefficient $\left(1\over 2F^2\right)s_{\beta_1\beta_3}$, we rearrange \eqref{6pt-1} by absorbing the term proportional to $\left(p_{\beta_1}+p_{\beta_2}+p_{\beta_3}\right)^2$ into ${\cal U}^{(2,3)}_1$, the left hand side of \eqref{6pt-1} becomes
\bea
\Sl_{\sigma\in OP(\{\alpha_1,\alpha_2\}\bigcup\{\beta_1,\beta_2,\beta_3\})}J(\sigma)={\cal V}^{(2,3)}J(\alpha_1)J(\alpha_2)J(\beta_1)J(\beta_2)J(\beta_3)+\left(1\over 2F^2\right)J(\alpha_1)J(\alpha_2)J(\beta_1,\beta_2,\beta_3),\Label{6pt-2}
\eea
where
\bea
{\cal V}^{(2,3)}\equiv {\cal U}^{(2,3)}_1+a_1^{(1,2)}\left(1\over 2F^2\right)^2{1\over p_1^2}s_{\beta_1\beta_3}\Label{V(2,3)}
\eea
which is shown by Fig. (\ref{redefining_coefficients_6pt}). The new defined coefficient of $(2,1)$ division is what we want. We need to prove ${\cal V}^{(2,3)}=\left(1\over 2F^2\right)^2$ for the $(2,3)$ division. In the next subsection, we have a general proof of this.

Let us consider the off-shell extension of the first term on the right hand side of \eqref{6pt-2},  assuming that we have already proved ${\cal V}^{(2,3)}=\left(1\over 2F^2\right)^2$. If all the $\alpha$'s and $\beta$'s are allowed to be off-shell, we replace $J(\alpha_i)$ by $J(A_i)$ and $J(\beta_i)$ by $J(B_i)$. Recalling that the coefficient of the off-shell extension should return to ${\cal V}^{(2,3)}=\left(1\over 2F^2\right)^2$ under the replacement $J(A_i)\to \alpha_i, J(B_i)\to \beta_i$ and  ${\cal V}^{(2,3)}$ can only be of the form ${1\over p_1^2}\Sl_{ij}c_{ij}p_i\cdot p_j$, the off-shell extension of ${\cal V}^{(2,3)}$ must have the form (see Fig. \ref{6pt-off-shell-extension1})
\bea
{\cal W}^{(2,3)}\equiv\left(1\over 2F^2\right)^2\left[1+{1\over p_1^2}\left(\Sl_{i=1}^2a_i^{(2,3)}p_A^2+\Sl_{i=1}^3b_i^{(2,3)}p_B^2\right)\right].\Label{W(2,3)}
\eea
Following a parallel discussion, we can do the same on the six-point relation with only one $\alpha$ and extend the coefficient of $(1,4)$ division to off-shell case (see Fig. \ref{6pt-off-shell-extension2})
\bea
{\cal W}^{(1,4)}\equiv\left(1\over 2F^2\right)^2{1\over p_1^2}\left(a_1^{(1,4)}p_A^2+\Sl_{i=1}^4b_i^{(1,4)}p_B^2\right).\Label{W(1,4)}
\eea

{\bf Eight-point example:} We now consider an eight-point example with three $\alpha$'s and five $\beta$'s. The formula of this example is given by \eqref{8pt-gen-U(1)-example} in section 3.

{\bf Step-1} To prove the eight-point example, we first express the left hand side of \eqref{8pt-gen-U(1)-example} by Berends-Giele recursion and then collect the contributions to a substructure of generalized $U(1)$-identity together. After applying generalized $U(1)$-identity, we get
\bea
&&\Sl_{\sigma\in OP(\{\alpha_1,\alpha_2,\alpha_3\}\bigcup\{\beta_1,\beta_2,\beta_3,\beta_4\})}J(\sigma)\nn
&=&\mathcal{U}^{(3,4)}_1J(\alpha_1)J(\alpha_2)J(\alpha_3)J(\beta_1)J(\beta_2)J(\beta_3)J(\beta_4)+\mathcal{U}^{(3,4)}_2J(\alpha_1)J(\alpha_2)J(\alpha_3)J(\beta_1,\beta_2,\beta_3)J(\beta_4)\nn
&&+\mathcal{U}^{(3,4)}_3J(\alpha_1)J(\alpha_2)J(\alpha_3)J(\beta_1)J(\beta_2,\beta_3,\beta_4)+\mathcal{U}^{(3,4)}_4J(\alpha_1,\alpha_2,\alpha_3)J(\beta_1)J(\beta_2)J(\beta_3)J(\beta_4)\nn
&&+\mathcal{U}^{(3,4)}_5J(\alpha_1,\alpha_2,\alpha_3)J(\beta_1,\beta_2,\beta_3)J(\beta_4)+\mathcal{U}^{(3,4)}_6J(\alpha_1,\alpha_2,\alpha_3)J(\beta_1)J(\beta_2,\beta_3,\beta_4).\Label{8pt-1}
\eea

 Again, we start from the $R+S=3$ divisions, there are two cases corresponding to the last two terms of the above equation. These cases only get contributions from diagrams with the off-shell leg $1$ connected to a four-point vertex. As shown in the six-point example, the coefficients ${\cal U}_5$ and ${\cal U}_6$ can be given by the off-shell extension ${\cal W}^{(1,2)}$ (Fig. \ref{4pt-off-shell-extension}), particularly
\bea
{\cal U}^{(3,4)}_5=\left(1\over 2F^2\right){1\over p_1^2}\left[p_1^2+a^{(1,2)} _1\left(p_{\alpha_1}+p_{\alpha_2}+p_{\alpha_3}\right)^2+b^{(1,2)}_1\left(p_{\beta_1}+p_{\beta_2}+p_{\beta_3}\right)^2\right]
\eea
and
\bea
{\cal U}^{(3,4)}_6=\left(1\over 2F^2\right){1\over p_1^2}\left[p_1^2+a^{(1,2)}_1\left(p_{\alpha_1}+p_{\alpha_2}+p_{\alpha_3}\right)^2+b^{(1,2)}_2\left(p_{\beta_2}+p_{\beta_3}+p_{\beta_4}\right)^2\right].
\eea

{\bf Step-2} The term $\left(p_{\alpha_1}+p_{\alpha_2}+p_{\alpha_3}\right)^2$ in ${\cal U}^{(3,4)}_5$ and ${\cal U}^{(3,4)}_6$ reduces $J(\alpha_1,\alpha_2,\alpha_3)$ to $J(\alpha_1)J(\alpha_2)J(\alpha_3)$ with a factor $\left(1\over 2F^2\right)s_{\alpha_1\alpha_3}$,  the term $\left(p_{\beta_1}+p_{\beta_2}+p_{\beta_3}\right)^2$ in ${\cal U}^{(3,4)}_5$ reduces $J(\beta_1,\beta_2,\beta_3)$ to $J(\beta_1)J(\beta_2)J(\beta_3)$ with a factor $\left(1\over 2F^2\right)s_{\beta_1\beta_3}$, while the term $\left(p_{\beta_2}+p_{\beta_3}+p_{\beta_4}\right)^2$ in ${\cal U}^{(3,4)}_6$ reduces $J(\beta_2,\beta_3,\beta_4)$ to $J(\beta_2)J(\beta_3)J(\beta_4)$ with a factor $\left(1\over 2F^2\right)s_{\beta_2\beta_4}$. As in the four-point example, we can redefine the coefficients so that \eqref{8pt-1} becomes
\bea
&&\Sl_{\sigma\in OP(\{\alpha_1,\alpha_2,\alpha_3\}\bigcup\{\beta_1,\beta_2,\beta_3,\beta_4\})}J(\sigma)\nn
&=&\mathcal{U}^{(3,4)}_1J(\alpha_1)J(\alpha_2)J(\alpha_3)J(\beta_1)J(\beta_2)J(\beta_3)J(\beta_4)+\mathcal{U}'^{(3,4)}_2J(\alpha_1)J(\alpha_2)J(\alpha_3)J(\beta_1,\beta_2,\beta_3)J(\beta_4)\nn
&&+\mathcal{U}'^{(3,4)}_3J(\alpha_1)J(\alpha_2)J(\alpha_3)J(\beta_1)J(\beta_2,\beta_3,\beta_4)+\mathcal{U}'^{(3,4)}_4J(\alpha_1,\alpha_2,\alpha_3)J(\beta_1)J(\beta_2)J(\beta_3)J(\beta_4)\nn
&&+\left(1\over 2F^2\right)J(\alpha_1,\alpha_2,\alpha_3)J(\beta_1,\beta_2,\beta_3)J(\beta_4)+\left(1\over 2F^2\right)J(\alpha_1,\alpha_2,\alpha_3)J(\beta_1)J(\beta_2,\beta_3,\beta_4),\Label{8pt-2}
\eea
where
\bea
{\cal U}'^{(3,4)}_2&=&\mathcal{U}^{(3,4)}_2+\left(1\over 2F^2\right)^2 a^{(1,2)}_1{1\over p_1^2}s_{\alpha_1\alpha_3},\nn
{\cal U}'^{(3,4)}_3&=&{\cal U}^{(3,4)}_3+\left(1\over 2F^2\right)^2a^{(1,2)}_1{1\over p_1^2}s_{\alpha_1\alpha_3},\nn
{\cal U}'^{(3,4)}_4&=&\mathcal{U}^{(3,4)}_4+\left(1\over 2F^2\right)^2{1\over p_1^2}\left[b^{(1,2)}_1s_{\beta_1\beta_3}+b^{(1,2)}_2s_{\beta_2\beta_4}\right].
\eea
When all the subcurrents go on-shell, the redefined coefficients ${\cal U}'^{(3,4)}_2$ and ${\cal U}'^{(3,4)}_3$ have the same pattern with ${\cal V}^{(3,4)}$ (by exchanging the roles of $\alpha$'s and $\beta$'s) in the six-point example, while ${\cal U}'^{(3,4)}_4$ has the same pattern with ${\cal V}^{(1,5)}$ in the six-point example. For instance, if we consider ${\cal U}'^{(3,4)}_2$
\begin{itemize}
\item the coefficients $\mathcal{U}^{(3,4)}_2$ get contributions from
\begin{itemize}
\item a) the diagrams with the off-shell leg $1$ connected to six-point vertices whose other legs are attached to subcurrents containing only $\alpha$ or $\beta$ elements (as shown in Fig. \ref{Two classes of diagrams.} (B))
\item b) the diagrams with the off-shell leg $1$ connected to four-point vertices, which contain substructures of generalized $U(1)$ identity (as shown in Fig. \ref{Two classes of diagrams.} (A) ).
\end{itemize}
Both cases has correspondence in the $\mathcal{U}_1^{(2,3)}$ of \eqref{V(2,3)} (with exchanging the roles of $\alpha$'s and $\beta$'s) and they have the same pattern with \eqref{V(2,3)} when the off-shell subcurrents goes on-shell.
\item The part $\left(1\over 2F^2\right)^2 a^{(1,2)}_1{1\over p_1^2}s_{\alpha_1\alpha_3}$ is same with $a_1^{(1,2)}\left(1\over 2F^2\right)^2{1\over p_1^2}s_{\beta_1\beta_3}$ in \eqref{V(2,3)} when exchanging the roles of $\alpha$'s and $\beta$'s.
\end{itemize}
Therefore, we can use the off-shell extensions \eqref{W(2,3)} and \eqref{W(1,4)} corresponding to ${\cal V}^{(2,3)}$ and ${\cal V}^{(1,4)}$ in the six-point example
\bea
{\cal U}'^{(3,4)}_2&=&\left(1\over 2F^2\right)^2{1\over p_1^2}\left[p_1^2+a^{(2,3)}_1\left(p_{\beta_1}+p_{\beta_2}+p_{\beta_3}\right)^2\right],\nn
 {\cal U}'^{(3,4)}_3&=&\left(1\over 2F^2\right)^2{1\over p_1^2}\left[p_1^2+a^{(2,3)}_2\left(p_{\beta_2}+p_{\beta_3}+p_{\beta_4}\right)^2\right],\nn
{\cal U}'^{(3,4)}_4&=&\left(1\over 2F^2\right)^2{1\over p_1^2}a_1^{(1,4)}\left(p_{\alpha_1}+p_{\alpha_2}+p_{\alpha_3}\right)^2.
\eea
%

%%%
\begin{figure}
  \centering
 \includegraphics[width=0.9\textwidth]{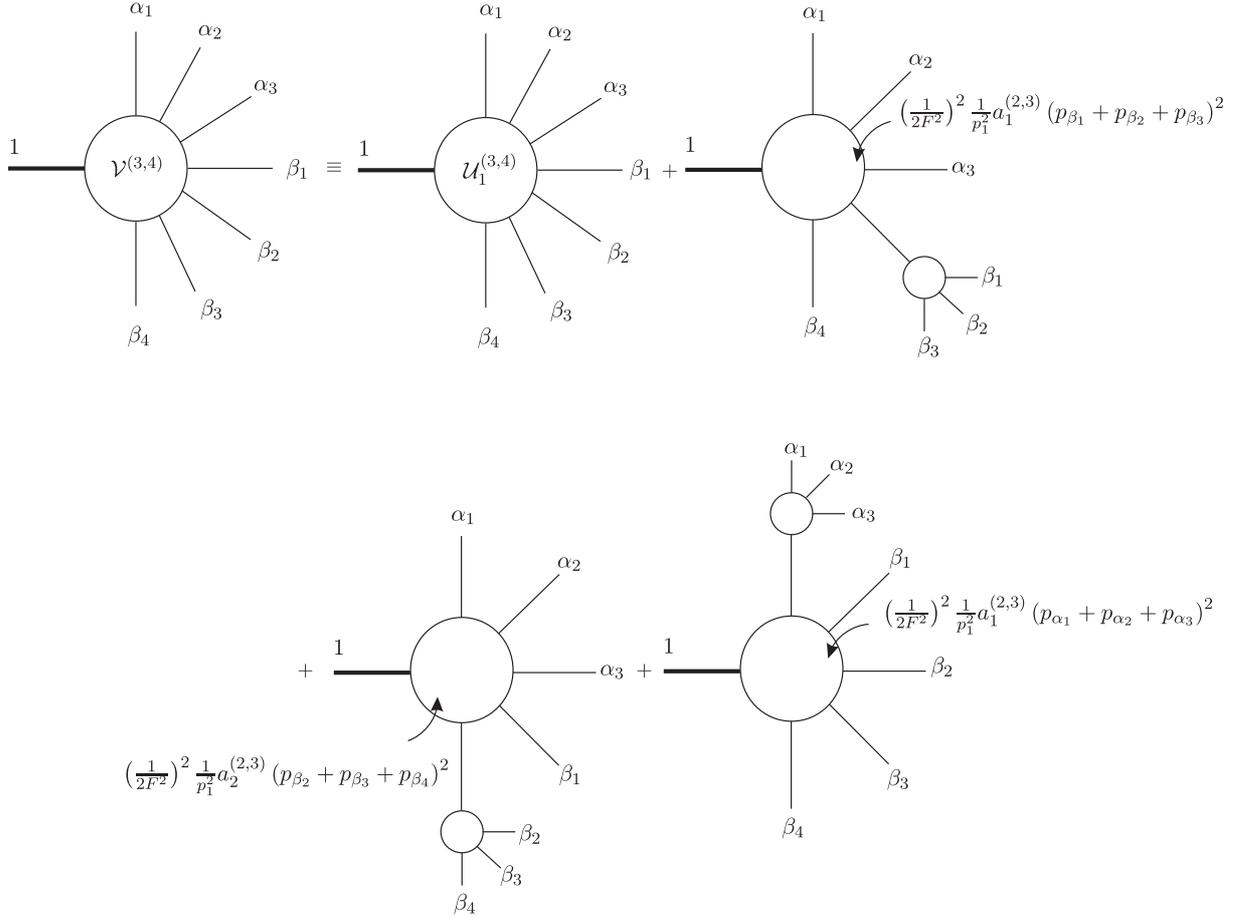}
 \caption{Redefinition of the coefficient of $(3,4)$ division for the identity with three $\alpha$'s and four $\beta$'s.}\label{redefining_coefficients_8pt}
\end{figure}
%%%

{\bf Step-3} Now we notice that $\left(p_{\beta_1}+p_{\beta_2}+p_{\beta_3}\right)^2$ in ${\cal U}'^{(3,4)}_2$ reduces $J(\beta_1,\beta_2,\beta_3)$ to $J(\beta_1)J(\beta_2)J(\beta_3)$ with a factor $\left(1\over 2F^2\right)s_{\beta_1\beta_3}$. Thus this term contributes to the $(3,4)$-division.
Similarly, the term $\left(p_{\beta_2}+p_{\beta_3}+p_{\beta_4}\right)^2$ in ${\cal U}'^{(3,4)}_3$ and
$\left(p_{\alpha_1}+p_{\alpha_2}+p_{\alpha_3}\right)^2$ in ${\cal U}'^{(3,4)}_4$ reduce $J(\beta_2,\beta_3,\beta_4)$ and $J(\alpha_1,\alpha_2,\alpha_3)$ to
$J(\beta_2)J(\beta_3)J(\beta_4)$ and $J(\alpha_1)J(\alpha_2)J(\alpha_3)$ respectively.
 Then, we can rearrange \eqref{8pt-2} again as (Fig. \eqref{redefining_coefficients_8pt})
\bea
&&\Sl_{\sigma\in OP(\{\alpha_1,\alpha_2,\alpha_3\}\bigcup\{\beta_1,\beta_2,\beta_3,\beta_4\})}J(\sigma)\nn
&=&\mathcal{V}^{(3,4)}J(\alpha_1)J(\alpha_2)J(\alpha_3)J(\beta_1)J(\beta_2)J(\beta_3)J(\beta_4)+\left(1\over 2F^2\right)^2J(\alpha_1)J(\alpha_2)J(\alpha_3)J(\beta_1,\beta_2,\beta_3)J(\beta_4)\nn
&&+\left(1\over 2F^2\right)^2J(\alpha_1)J(\alpha_2)J(\alpha_3)J(\beta_1)J(\beta_2,\beta_3,\beta_4)+\left(1\over 2F^2\right)^2J(\alpha_1,\alpha_2,\alpha_3)J(\beta_1)J(\beta_2)J(\beta_3)J(\beta_4)\nn
&&+\left(1\over 2F^2\right)J(\alpha_1,\alpha_2,\alpha_3)J(\beta_1,\beta_2,\beta_3)J(\beta_4)+\left(1\over 2F^2\right)J(\alpha_1,\alpha_2,\alpha_3)J(\beta_1)J(\beta_2,\beta_3,\beta_4),\Label{8pt-3}
\eea
where
\bea
\mathcal{V}^{(3,4)}&=&{\cal U}^{(3,4)}_1+\left(1\over 2F^2\right)^2{1\over p_1^2}\left[a^{(2,3)}_1 s_{\beta_1\beta_3}+a^{(2,3)}_2s_{\beta_2\beta_4}+a_1^{(1,4)}s_{\alpha_1\alpha_3}\right].
\eea
Thus we only need to prove $\mathcal{V}^{(3,4)}=\left(1\over 2F^2\right)^3$. We leave the proof to the next subsection.

{\bf General Discussion:} In general, when we consider the generalized $U(1)$-identity \eqref{off-shell-gen-U(1)} with $r$ $\alpha$'s and $s$ $\beta$'s, we can use Berends-Giele recursion to rewrite the left hand side and collect terms corresponding to a same substructure as shown in Fig. \ref{Two classes of diagrams.}. Applying the lower-point identity to the substructures and summing the coefficients for any given division, we reexpress the left hand side of \eqref{off-shell-gen-U(1)} by \eqref{off-shell-gen-U(1)-0} or briefly by
\bea
\Sl_{D}{\cal U}^{(r,s)}_DJ(A_1)\dots J(A_{R_D})J(B_1)\dots J(B_{S_D}).\Label{off-shell-gen-U(1)-2}
\eea

We start from the divisions with $R_D+S_D=3$, i.e., $(1,2)$ division and $(2,1)$ division. The contributing diagrams are those in the four-point example with replacing the on-shell lines by off-shell currents. Thus it has the form of the off-shell extension \eqref{W(1,2)}. Since $p_{A_i}^2J_{A_i}$ and $p_{B_i}^2J_{B_i}$ in \eqref{W(1,2)} will further reproduce  divisions of $A_i$ and $B_i$ with coefficients $\Sl{1\over p_1^2}c_{ij}p_i\cdot p_j$, we absorb all these contributions into the corresponding divisions with $R_D+S_D>3$. The only left contribution for divisions with $R_D+S_D=3$ is the first term of \eqref{W(1,2)} which gives rise to the expected coefficients.

Then we consider divisions with $R_D+S_D=5$, which get both contributions from its corresponding ${\cal U}_D^{(r,s)}$ in \eqref{off-shell-gen-U(1)-2} as well as $p_{A_i}^2J_{A_i}$ and $p_{B_i}^2J_{B_i}$ in the off-shell extension  \eqref{W(1,2)} of four-point case. Since the coefficients for divisions with $R_D+S_D=5$ are defined in the same way with the $\mathcal{V}^{(r',s')}$ $(r'+s'=5)$ in the six-point example, they are just the off-shell extensions \eqref{W(2,3)}, and  \eqref{W(1,4)}. Again, the terms containing $p_{A_i}^2J_{A_I}$ and $p_{B_i}^2J_{B_i}$ in \eqref{W(2,3)} and \eqref{W(1,4)} are absorbed into the divisions with $R_D+S_D>5$. The left contributions are those expected coefficients for divisions with $R_D+S_D=5$.

 Redefining the coefficients level by level, we finally have \eqref{off-shell-gen-U(1)-1} where all the coefficients of $R+S<r+s$ divisions match with those in the final formula of the identity \eqref{off-shell-gen-U(1)}. The coefficient  ${\cal V}^{(r,s)}$ defined by this method only get contributions from the ${\cal U}^{(r,s)}_1$ corresponding to the $(r,s)$ division  as well as the off-shell extensions of ${\cal V}^{(R,S)}$ with $R+S<r+s$. Both cases contain terms proportional to ${1\over p_1^2}s_{ij}$, where $i$ and $j$ denote arbitrary external on-shell lines. Thus ${\cal V}^{(r,s)}$ has the general form
\bea
{\cal V}^{(r,s)}={1\over p_1^2}\left(\Sl_{1\leq i< j\leq r}c_{\alpha_i\alpha_j}s_{\alpha_i\alpha_j}+\Sl_{1\leq i< j\leq s}c_{\beta_i\beta_j}s_{\beta_i\beta_j}+\Sl_{i=1}^r\Sl_{j=1}^sc_{\alpha_i\beta_j}s_{\alpha_i\beta_j}\right).~~\Label{gen-V(r,s)}
\eea
In the remaining part of this section, we will solve the coefficients $c$'s to show that ${\cal V}^{(r,s)}$ has the expected form.

%%%%%%%%%%%%%%%%%%%%%%%%%%%%%%%%%%%%%%%%%%%%%%%%%%%%%%%%%%%%%%%%%%%%%%%%%%%%%%%%
\subsection{Solving  ${\cal V}^{(r,s)}$}
%%%%%%%%%%%%%%%%%%%%%%%%%%%%%%%%%%%%%%%%%%%%%%%%%%%%%%%%%%%%%%%%%%%%%%%%%%%%%%%%
In the above discussion, we have shown that the left hand side of the generalized $U(1)$-identity \eqref{off-shell-gen-U(1)} could be rearranged into the form \eqref{off-shell-gen-U(1)-1}. All the coefficients of divisions in \eqref{off-shell-gen-U(1)-1} with $R_D+S_D<r+s$ are those on the right hand side of the identity \eqref{off-shell-gen-U(1)}.
Only the coefficient ${\cal V}^{(r,s)}$ for $(r,s)$-division are undetermined. Now let us prove that ${\cal V}^{(r,s)}$ has the right form, i.e.,
\bea
{\cal V}^{(r,s)}=\left(1\over 2F^2\right)^{r+s-1\over 2 }\delta(|r-s|-1).~~\Label{V(r,s)}
\eea
%
%%%%%%%%%%%%
\subsubsection{$r=1$}
%%%%%%%%%%%%%%%%
\begin{figure}
  \centering
 \includegraphics[width=0.75\textwidth]{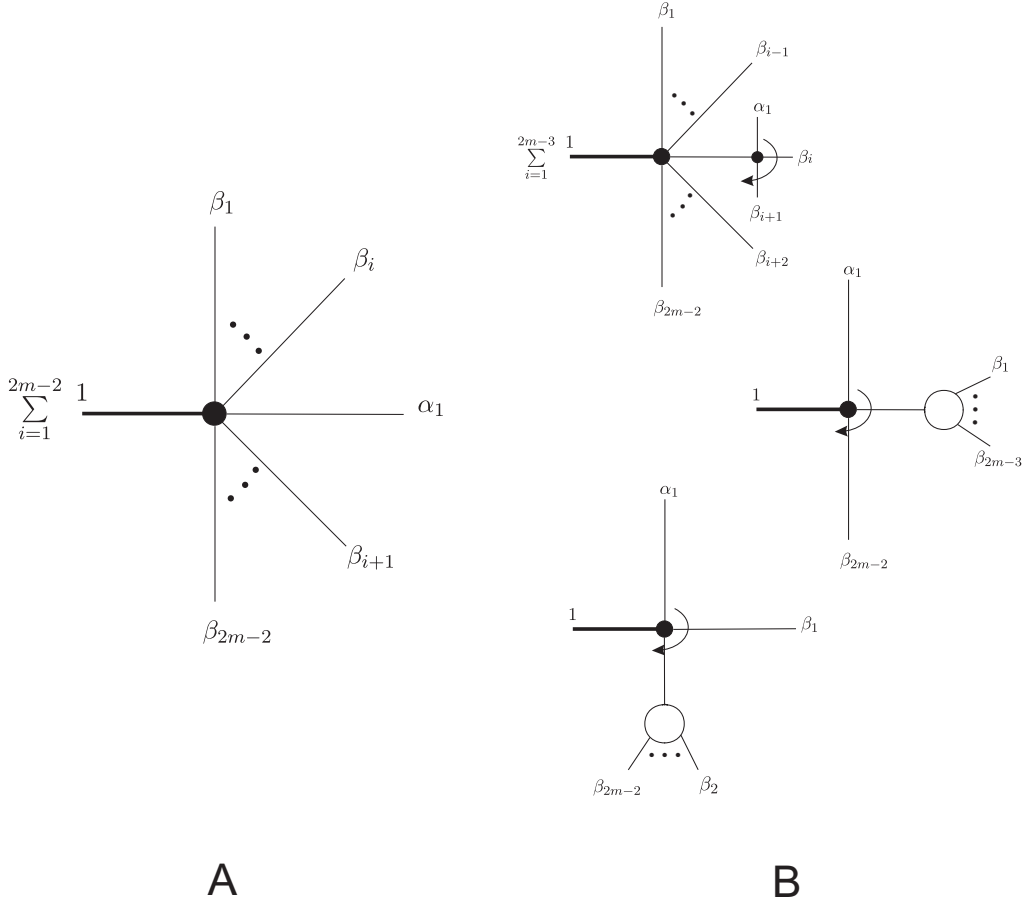}
 \caption{Diagrams contributing to ${\cal V}^{(1,2m-2)}$. A curved arrow line denotes the sum over all the possible three positions of $\alpha_1$ around the four-point vertices.}\label{figU1}
\end{figure}
%%%
In this case, \eqref{off-shell-gen-U(1)} becomes the $U(1)$-identity for $2m=s+2$-point currents, which has been studied in \cite{Chen:2013fya}.
The ${\cal V}^{(1,2)}$ for four-point relation with one $\alpha$ and two $\beta$'s is
\bea
{\cal V}^{(1,2)}={1\over 2F^2}.
\eea
The coefficient ${\cal V}^{(1,4)}$ for six-point relation with only one $\alpha$ vanishes.
Generically, ${\cal V}^{(1,2m-2)}$ only gets contributions from the diagrams in Fig. \ref{figU1}. Following direct calculation which has been shown in \cite{Chen:2013fya}, we find that
\bea
{\cal V}^{(1,s)}=0, (\text{for $s>2$}).\Label{V(1,s)}
\eea
Hence ${\cal V}^{(1,2m-2)}$ satisfies the form \eqref{V(r,s)}.

\subsubsection{$r>1$}

To solve ${\cal V}^{(r,s)}$ for $r>1$, we consider the following combination of currents
\bea
I(\alpha_1\mid \alpha_2,\dots,\alpha_r;\beta_1,\dots,\beta_s)\equiv\Sl_{\rho\in OP(\{\alpha_2,\dots,\alpha_r\}\bigcup\{\beta_1,\dots,\beta_s\})}\left[\Sl_{\sigma\in OP(\{\alpha_1\}\bigcup\{\rho\})}J(\sigma)\right],\Label{I}
\eea
 where we have combined left hand sides of a $U(1)$-identity and a generalized $U(1)$-identity with $(r-1)$ $\alpha$'s. Now lets consider the coefficient
of the $(r,s)$-division of $I(\alpha_1|\alpha_2,\dots,\alpha_r;\beta_1,\dots,\beta_s)$.
This can be obtained in two different ways:
\begin{itemize}
\item [\bf(a)] For a given permutation $\rho\in OP(\{\alpha_2,\dots,\alpha_r\}\bigcup\{\beta_1,\dots,\beta_s\})$, we apply the $U(1)$-identity \eqref{off-shell-U(1)} with $\{\rho\}$ as the $\{\beta\}$ set.
    Then we have
    \bea
  \Sl_{\sigma\in OP({\alpha_1}\bigcup{\rho})}J(\sigma)=\Sl_{\{\rho\}\to\{\rho_L\}\{\rho_R\}}\left(1\over 2F^2\right)J(\alpha_1)J(\{\rho_L\})J(\{\rho_R\}).
    \eea
 Here we summed over divisions $\{\rho\}\to\{\rho_L\}\{\rho_R\}$ on the right hand side.
  Substituting above expression into \eqref{I} and rearranging the summations, we reexpress the combination $I(\alpha_1\mid\alpha_2,\dots,\alpha_r;\beta_1,\dots,\beta_s)$ by
   \bea
   &&I(\alpha_1\mid\alpha_2,\dots,\alpha_r;\beta_1,\dots,\beta_s)\nn
   &=&\Sl_{\rho\in OP(\{\alpha_2,\dots,\alpha_r\}\bigcup\{\beta_1,\dots,\beta_s\})}\Sl_{\{\rho\}\to\{\rho_L\}\{\rho_R\}}\left(1\over 2F^2\right)J(\alpha_1)J(\{\rho_L\})J(\{\rho_R\})\nn
   &=&\Sl_{\tiny
             \begin{array}{c}
              \{\alpha_2,\dots,\alpha_r\}\to\{\alpha_L\}\{\alpha_R\} \\
               \{\beta_1,\dots,\beta_s\}\to\{\beta_L\}\{\beta_R\} \\
             \end{array}
           }
 \left(1\over 2F^2\right)J(\alpha_1)\left[\Sl_{\rho_L\in OP(\{\alpha_L\}\bigcup\{\beta_L\})}J(\{\rho_L\})\Sl_{\rho_R\in OP(\{\alpha_R\}\bigcup\{\beta_R\})}J(\{\rho_R\})\right],\nn
   \eea
   where $\Sl_{\rho_L\in OP(\{\alpha_L\}\bigcup\{\beta_L\})}J(\{\rho_L\})$ and $\Sl_{\rho_R\in OP(\{\alpha_R\}\bigcup\{\beta_R\})}J(\{\rho_R\})$ are two lower-point substructures of generalized $U(1)$-identity \eqref{off-shell-U(1)}. From recursive assumption, we know that both the coefficient  ${\cal V}^{(r_L,s_L)}$ for the $(r_L,s_L)$-substructure and the coefficient ${\cal V}^{(r_R,s_R)}$ for the $(r_R,s_R)$-substructure satisfy \eqref{V(r,s)}.
   Thus the coefficient ${\cal V}_I^{(r,s)}$ of the $(r,s)$ division of $I(\alpha_1\mid \alpha_2,\dots,\alpha_r;\beta_1,\dots,\beta_s)$ is
    \bea
   {\cal V}_I^{(r,s)}&=&\Sl_{\tiny
             \begin{array}{c}
              \{\alpha_2,\dots,\alpha_r\}\to\{\alpha_L\}\{\alpha_R\} \\
               \{\beta_1,\dots,\beta_s\}\to\{\beta_L\}\{\beta_R\} \\
             \end{array}
           }\left(1\over 2F^2\right){\cal V}^{(r_L,s_L)}{\cal V}^{(r_R,s_R)}\nn
   &=&\Sl_{\tiny
             \begin{array}{c}
              \{\alpha_2,\dots,\alpha_r\}\to\{\alpha_L\}\{\alpha_R\} \\
               \{\beta_1,\dots,\beta_s\}\to\{\beta_L\}\{\beta_R\} \\
             \end{array}
           }\left(1\over 2F^2\right)^{{r_L+s_L-1\over 2}+{r_R+s_R-1\over 2}+1}\delta(|r_L-s_L|-1)\delta(|r_R-s_R|-1).\Label{V-I}\nn
    \eea
   The delta functions impose constraints on $r=r_L+r_R-1$ and $s=s_L+s_R$. The only nonzero contributions are the cases with $r$, $s$ satisfying
   \bea
    r=s-1,r=s+1, r=s+3.
   \eea
  \emph{i)} For $r=s-1$, only  terms with $r_L=s_L-1$ and $r_R=s_R-1$ in \eqref{V-I} are nonzero. Thus we have
  \bea
  {\cal V}_I^{(s-1,s)}&=&\Sl_{s_L=1
           }^{s-1}\left(1\over 2F^2\right)^{{(s_L-1)+s_L-1\over 2}+{(s-s_L-1)+(s-s_L)-1\over 2}+1}=\Sl_{s_L=1
           }^{s-1}\left(1\over 2F^2\right)^{s-1}=\left(1\over 2F^2\right)^{{r+s-1\over 2}}(s-1).\Label{V-I1}\nn
  \eea
 \emph{ii)} For $r=s+1$, both terms with $r_L=s_L-1$, $r_R=s_R+1$ and terms with $r_L=s_L+1$, $r_R=s_R-1$ contribute. Then
 \bea
 {\cal V}_I^{(s+1,s)}&=&\Sl_{s_L=1
           }^{s-1}\left(1\over 2F^2\right)^{{(s_L-1)+s_L-1\over 2}+{(s-s_L+1)+(s-s_L)-1\over 2}+1}+\Sl_{s_L=0
           }^{s}\left(1\over 2F^2\right)^{{(s_L-1)+s_L-1\over 2}+{(s-s_L+1)+(s-s_L)-1\over 2}+1}\nn
  &=&\left(1\over 2F^2\right)^{{r+s-1\over 2}}2s.~~\Label{V-I2}
  \eea
 \emph{iii)} For  $r=s+3$, the nonvanishing terms are those with $r_L=s_L+1$, $r_R=s_R+1$. Thus we get
  \bea
  {\cal V}_I^{(s+3,s)}=\Sl_{s_L=0
           }^{s}\left(1\over 2F^2\right)^{{(s_L+1)+s_L-1\over 2}+{(s-s_L+1)+(s-s_L)-1\over 2}+1}=\left(1\over 2F^2\right)^{{r+s-1\over 2}}(s+1).~~\Label{V-I3}
  \eea
\item [\bf(b)] The combination of currents $I(\alpha_1\mid\alpha_2,\dots,\alpha_r;\beta_1,\dots,\beta_s)$ can be expressed from another angle:
Considering a given  $\{\rho\}\in OP(\{\alpha_1\}\bigcup\{\alpha_2,\dots,\alpha_r\})$ as  the $\{\alpha\}$ set on the left hand side of \eqref{off-shell-gen-U(1)}, we have a combination of currents $\Sl_{\{\sigma\}\in OP(\{\rho\}\bigcup\{\beta_1,\dots,\beta_s\})}J(\sigma)$. After summing over all
$\{\rho\}\in OP(\{\alpha_1\}\bigcup\{\alpha_2,\dots,\alpha_r\})$, we express $I(\alpha_1\mid \alpha_2,\dots,\alpha_r;\beta_1,\dots,\beta_s)$ by
\bea
I(\alpha_1\mid \alpha_2,\dots,\alpha_r;\beta_1,\dots,\beta_s)=\Sl_{\{\rho\}\in OP(\{\alpha_1\}\bigcup\{\alpha_2,\dots,\alpha_r\})}\left[\Sl_{\{\sigma\}\in OP(\{\rho\}\bigcup\{\beta_1,\dots,\beta_s\})}J(\sigma)\right].\Label{I-1}
\eea
Expressing each $\Sl_{\{\sigma\}\in OP(\{\rho\}\bigcup\{\beta_1,\dots,\beta_s\})}J(\sigma)$ by \eqref{off-shell-gen-U(1)-1}, we collect the coefficients of $(r,s)$ division for $I(\alpha_1\mid \alpha_2,\dots,\alpha_r;\beta_1,\dots,\beta_s)$. There are two parts of contributions $\mathcal{A}^{(r,s)}$ and $\mathcal{B}^{(r,s)}$:
\begin{itemize}
\item [\emph{i)}] the first part $\mathcal{A}^{(r,s)}$ is the sum of the
${\cal V}^{(r,s)}$ coefficients for all possible $\rho\in OP(\{\alpha_1\}\bigcup\{\alpha_2,\dots,\alpha_r\})$,
\item [\emph{ii)}] the second part $\mathcal{B}^{(r,s)}$ is the sum of terms with $(r-2,s)$ divisions containing a nontrivial subcurrent $J(\phi\in OP(\{\alpha_1\}\bigcup\{\alpha_i,\alpha_{i+1}\}))$.
\end{itemize}

 As shown in the previous subsection, the terms in $\mathcal{B}^{(r,s)}$ already have the expected coefficients $\left(1\over 2F^2\right)^{(r-2)+s-1\over 2}\delta(|r-2-s|-1)$. Collecting the $(r-2,s)$ divisions containing subcurrents $J(\alpha_1,\alpha_i,\alpha_{i+1})$, $J(\alpha_i,\alpha_1,\alpha_{i+1})$, $J(\alpha_i,\alpha_{i+1},\alpha_1)$ and applying the $U(1)$-identity with one $\alpha$ and two $\beta$'s, we obtain a term with $(r,s)$ division for $I(\alpha_1\mid \alpha_2,\dots,\alpha_r;\beta_1,\dots,\beta_s)$. The coefficient is
\bea
\left(1\over 2F^2\right)^{{(r-2)+s-1\over 2}+1}\delta(|r-2-s|-1)=\left(1\over 2F^2\right)^{{r+s-1\over 2}}\delta(|r-2-s|-1).
\eea
After summing over $i=2,3,\dots,r-1$, we get ${\cal B}^{(r,s)}$
\bea
{\cal B}^{(r,s)}=\Sl_{i=2}^{r-1}\left(1\over 2F^2\right)^{{r+s-1\over 2}}\delta(|r-2-s|-1).
\eea
Therefore, ${\cal V}_I^{(r,s)}$ is given by
\bea
{\cal V}_I^{(r,s)}={\cal A}^{(r,s)}+{\cal B}^{(r,s)}={\cal A}^{(r,s)}+\left(1\over 2F^2\right)^{{r+s-1\over 2}}(r-2)\delta(|r-2-s|-1).
\eea
Again, the delta function imposes a constraint on $r$ and $s$. The only nonzero cases are $r=s+1$ and $r=s+3$.

\emph{i)} For $r=s+1$, we have
\bea
{\cal V}_I^{(s+1,s)}={\cal A}^{(s+1,s)}+\left(1\over 2F^2\right)^{{r+s-1\over 2}}(s-1).
\eea
\emph{ii)} For $r=s+3$, we have
\bea
{\cal V}_I^{(s+3,s)}={\cal A}^{(s+3,s)}+\left(1\over 2F^2\right)^{{r+s-1\over 2}}(s+1).
\eea
\end{itemize}
Comparing these expressions of ${\cal V}_I^{(r,s)}$ derived from ${\bf(a)}$ approach with those from ${\bf(b)}$ approach, we immediately conclude that
\bea
{\cal A}^{(r,s)}=\left(1\over 2F^2\right)^{{r+s-1\over 2}}r\delta(|r-s|-1).~~\Label{A(r,s)}
\eea
Then ${\cal A}^{(r,s)}$ can be expanded as
\bea
{\cal A}^{(r,s)}=\left(1\over 2F^2\right)^{{r+s-1\over 2}}{1\over p_1^2}\left(\Sl_{1\leq i<j\leq r}d_{\alpha_i\alpha_j}s_{\alpha_i\alpha_j}+\Sl_{1\leq i<j\leq s}d_{\beta_i\beta_j}s_{\beta_i\beta_j}+\Sl_{i=1}^r\Sl_{j=1}^sd_{\alpha_i\beta_j}s_{\alpha_i\beta_j}\right),~~\Label{A(r,s)-1}
\eea
where  momentum conservation and on-shell conditions have  been used; $d_{ij}$ ($i$, $j$ can be any $\{\alpha\}$ or $\{\beta\}$ elements) are defined by
\bea
d_{\alpha_i\alpha_j}=d_{\beta_i\beta_j}=d_{\alpha_i\beta_j}=r\delta(|r-s|-1).\Label{d0}
\eea

If we exchange the roles of $\{\alpha\}$ and $\{\beta\}$ in $I(\alpha_1\mid \alpha_2,\dots,\alpha_r;\beta_1,\dots,\beta_s)$, we get another combination of currents
\bea
I(\alpha_1,\dots,\alpha_r;\beta_1\mid \beta_2,\dots,\beta_s)\equiv\Sl_{\rho\in OP(\{\alpha_1,\dots,\alpha_r\}\bigcup\{\beta_2,\dots,\beta_s\})}\left[\Sl_{\sigma\in OP(\{\rho\}\bigcup\{\beta_1\})}J(\sigma)\right],\Label{I'}
\eea
which has an equivalent form
\bea
I(\alpha_1,\dots,\alpha_r;\beta_1\mid \beta_2,\dots,\beta_s)=\Sl_{\rho\in OP(\{\alpha_1,\dots,\alpha_r\}\bigcup\{\rho\})}\left[\Sl_{\sigma\in OP(\{\rho\}\bigcup\{\beta_1\})}J(\sigma)\right].\Label{I'-1}
\eea
Following a parallel discussion, we have
\bea
{\cal A'}^{(s,r)}=\left(1\over 2F^2\right)^{{r+s-1\over 2}}{1\over p_1^2}\left(\Sl_{1\leq i<j\leq r}d'_{\alpha_i\alpha_j}s_{\alpha_i\alpha_j}+\Sl_{1\leq i<j\leq s}d'_{\beta_i\beta_j}s_{\beta_i\beta_j}+\Sl_{i=1}^r\Sl_{j=1}^sd'_{\alpha_i\beta_j}s_{\alpha_i\beta_j}\right),~~\Label{A'(r,s)-1}
\eea
where ${\cal A'}^{(s,r)}$ is similar with ${\cal A}^{(r,s)}$ but defined from $I(\alpha_1,\dots,\alpha_r;\beta_1\mid \beta_2,\dots,\beta_s)$ instead. The coefficients $d'$s are given by
\bea
d'_{\alpha_i\alpha_j}=d'_{\beta_i\beta_j}=d'_{\alpha_i\beta_j}=s\delta(|s-r|-1).\Label{d'0}
\eea
%
%%%
\begin{figure}
  \centering
 \includegraphics[width=0.7\textwidth]{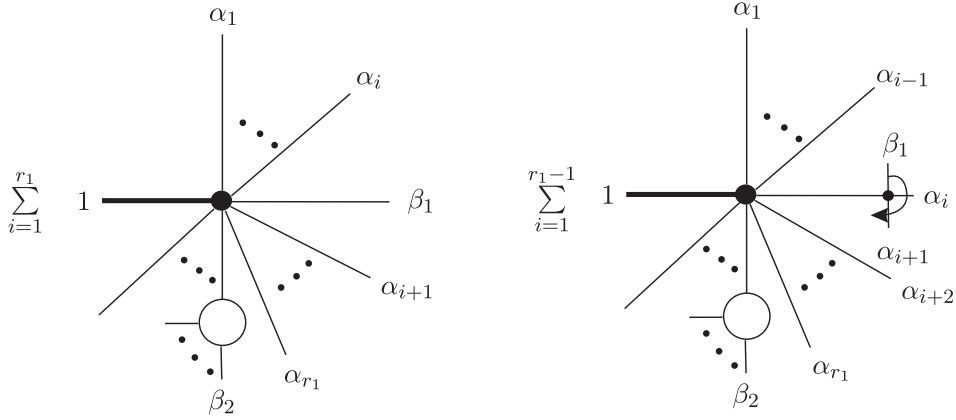}
 \caption{The diagrams cancel out with $\beta_1$, $\beta_2$ from reductions of different subcurrents when $r_1$ is even.}\label{Cancel1}
\end{figure}
%%%

%%%
\begin{figure}
  \centering
 \includegraphics[width=0.7\textwidth]{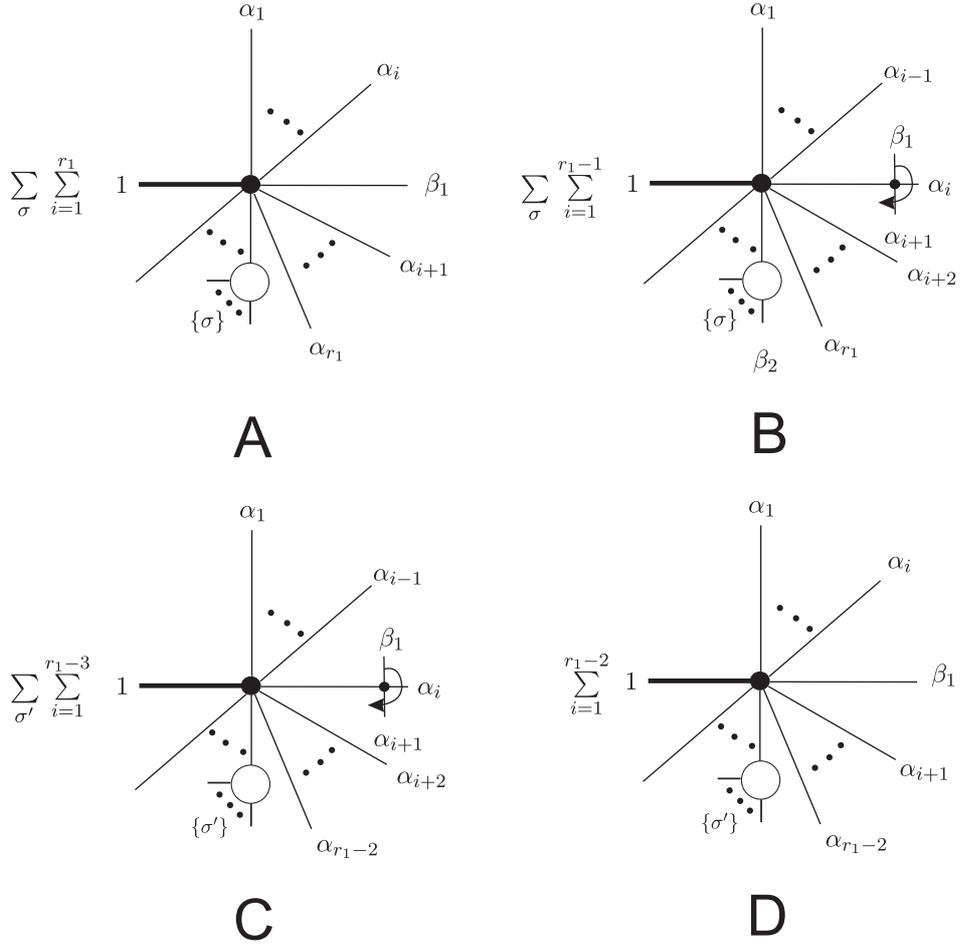}
 \caption{The diagrams cancel out with $\beta_1$, $\beta_2$ from reductions of different subcurrents when $r_1$ is odd. In A and B, we summed over $\sigma\in OP(\alpha_{r_1+1},\dots,\alpha_{r_1+s_1-1})\bigcup\{\beta_2,\dots,\beta_{s_1}\}$ for given $s_1$. In C and D, we summed over $\sigma'\in OP(\alpha_{r_1-1},\dots,\alpha_{r_1+s_1-1})\bigcup\{\beta_2,\dots,\beta_{s_1}\}$.}\label{Cancel2}
\end{figure}
%%%
Recalling that ${\cal A}^{(r,s)}$ is given by sum of the ${\cal V}^{(r,s)}$ corresponding to different permutations $\{\rho\}\in OP(\{\alpha_1\}\bigcup\{\alpha_2,\dots,\alpha_r\})$ in \eqref{I-1} and ${\cal V}^{(r,s)}$ have the general pattern \eqref{gen-V(r,s)}, we express ${\cal A}^{(r,s)}$ in \eqref{A(r,s)-1} by the general expression \eqref{gen-V(r,s)} of ${\cal V}^{(r,s)}$. Comparing the coefficients of each $s_{\alpha_i\beta_j}$ $s_{\alpha_i\alpha_j}$ and $s_{\beta_i\beta_j}$ on both sides of \eqref{A(r,s)-1}, we obtain a set of equations for $c_{ij}$ where either $i$ or $j$ can be $\alpha$ or $\beta$ elements. Similarly, when expressing ${\cal A'}^{(s,r)}$ by the ${\cal V}^{(r,s)}$'s corresponding to different permutations $\rho\in OP(\{\beta_1\}\bigcup\{\beta_2,\dots,\beta_s\})$ in \eqref{I'-1},  we can also establish the relations between $c_{ij}$ and $d'_{ij}$.
Let us solve the coefficients $c_{\alpha_i\beta_j}$, $c_{\alpha_i\alpha_j}$ and $c_{\beta_i\beta_j}$ from these equations.

\begin{itemize}
\item {\bf $c_{\alpha_i\beta_j}$}

We now solve $c_{\alpha_i\beta_j}$ from their relations with $d_{\alpha_i\beta_j}$ in \eqref{A(r,s)-1}. Noticing that any permutation $\rho$ in  the first sum of \eqref{I-1} has the general form $\{\alpha_2,\dots,\alpha_i,\alpha_1,\alpha_{i+1},\dots,\alpha_r\}$, the coefficient of $s_{\alpha_1\beta_j}$ in the second sum should be $c_{\alpha_i\beta_j}$. This is because the $\alpha_1$ in $\{\alpha_2,\dots,\alpha_i,\alpha_1,\alpha_{i+1},\dots,\alpha_r\}$ is inserted at the $i$-th position and plays as the $\alpha_i$ in the standard permutation $\{\alpha_1,\alpha_2,\alpha_3,\dots,\alpha_r\}$. Thus we get the following equation
\bea
d_{\alpha_1\beta_j}=\Sl_{i=1}^rc_{\alpha_i\beta_j}, (j=1,2,\dots,s).\Label{d1}
\eea
Similarly, the coefficient of $s_{\alpha_l\beta_j}$ $(2\leq l\leq r)$  in the sum over $\sigma\in OP(\{\alpha_2,\dots,\alpha_i,\alpha_1,\alpha_{i+1},\dots,\alpha_r\}\bigcup\{\beta\})$ in \eqref{I-1} for given $i$ is
\bea
\Biggl\{
  \begin{array}{cc}
    c_{\alpha_l\beta_j}& (i<l\leq r) \\
    c_{\alpha_{l-1}\beta_j} & (1<l\leq i-1) \\
  \end{array}.
\eea
Thus $d_{\alpha_i\beta_j}$ with $i=2,\dots,r$ is given by
\bea
d_{\alpha_i\beta_j}=(l-1)c_{\alpha_l\beta_j}+(r-l+1)c_{\alpha_{l-1}\beta_j}, (l=2,\dots,r, j=1\dots s).\Label{d2}
\eea

In the same way, when considering ${\cal A}'^{(s,r)}$ and the combination \eqref{I'}, we obtain the relations between $d'$'s and $c$'s
\bea
d'_{\alpha_i\beta_1}&=&\Sl_{j=1}^sc_{\alpha_i\beta_j}, (i=1,2,\dots,r)\Label{d'1}
\eea
\bea
d'_{\alpha_i\beta_j}&=&(k-1)c_{\alpha_i\beta_k}+(s-k+1)c_{\alpha_{i}\beta_{k-1}}, (i=1,\dots,r, k=2\dots s).\Label{d'2}
\eea

We first prove that $c_{\alpha_1\beta_1}=c_{\alpha_1\beta_2}$. Considering positions of $\beta_1$ and $\beta_2$, we can classify the contributing diagrams into two types:
\begin{itemize}
 \item[\emph{i)}]$ \beta_1$ and $\beta_2$ come from reduction of a same subcurrent.
\item[\emph{ii)}] $\beta_1$ and $\beta_2$ come from reduction of different subcurrents.
 \end{itemize}
The factors $c_{\alpha_1\beta_1}$ and $c_{\alpha_1\beta_2}$ receive equal contributions from the first types of diagrams.
For the second type, we can always find diagrams cancel with each other. To see this, we assume that the last $\alpha$ element in front of $\beta_2$ (or in the same substructure with $\beta$) is $\alpha_{r_1}$.
\begin{itemize}
\item [] If $r_1$ is even, the diagrams are typically given by Fig. \ref{Cancel1}.  The left diagram in Fig. \ref{Cancel1} contribute a $\frac{r_1}{2}s_{\alpha_1\beta_1}$\footnote{For convenience, we neglect a factor $\left(1\over 2F^2\right)^{{r+s-1\over 2}}$ in the remaining discussion and put the factor back in the final result.}, while the right diagram contribute a $-\frac{r_1}{2}s_{\alpha_1\beta_1}$. Thus these two contributions to $c_{\alpha_1\alpha_1}$ cancel out. Since the $r_1$ is even, there are odd number of legs in front of $\beta_2$. From Feynman rules, such diagrams do not contribute to $c_{\alpha_1\beta_2}$.
\item [] If $r_1$ is odd, the diagrams in Fig. \ref{Cancel2} should be taken into account. The A, B diagrams of Fig. \ref{Cancel2} contribute ${r_1-2\over 2}$ and  $-{r_1-2\over 2}$ to $c_{\alpha_1\beta_1}$, while the diagrams C and D of Fig. \ref{Cancel2} contribute  ${r_1-4\over 2}$ and  $-{r_1-4\over 2}$ to $c_{\alpha_1\beta_1}$. Thus $c_{\alpha_1\beta_1}$ does not get any nonzero contribution from Fig. \ref{Cancel2}.
    When considering $c_{\alpha_1\beta_2}$, we find that diagrams A, B, C, D in Fig. \ref{Cancel2} contribute $r_1$, $-(r_1-1)$, $r_1-3$, $-(r_1-2)$. Thus $c_{\alpha_1\beta_2}$ also does not get any nonzero contribution from these diagrams.
\end{itemize}

%For diagrams may only net contribute to $c_{\alpha_1\beta_1}$, we can always find a pair of diagrams cancel with each other. The two diagrams
%in figure \ref{Cancel1} provide an example, and $r_1$ must be even. The point is that one can always find a pair of diagrams. In one diagram $\beta_1$ is an on-shell leg, while in the other diagram $\beta_1$ is in a substructure with a sum over $OP(\{\alpha_i,\alpha_{i+1}\bigcup\{\beta_1\}\})$. The left diagram contribute a $\frac{r_1}{2}s_{\alpha_1\beta_1}$ while the right diagram contribute a $-\frac{r_1}{2}s_{\alpha_1\beta_1}$, thus these two contributions cancel out.
%Similarly, for diagrams may only contribute to $c_{\alpha_1\beta_2}$, we can always find four diagrams cancel together. The four diagrams are shown in figure \ref{Cancel2}. The four diagrams A, B, C, D's net contribution are $\frac{r_1+1}{2}$, $\frac{r_1-3}{2}$, $-\frac{r_1-1}{2}$, $-\frac{r_1-1}{2}$. Thus these four contributions cancel out.

Now we substitute $c_{\alpha_1\beta_1}=c_{\alpha_1\beta_2}$ into \eqref{d'2} with $i=1, k=2$ and remember $d'$ have the form \eqref{d'0} we have
\bea
c_{\alpha_1\beta_1}=c_{\alpha_1\beta_2}=\delta(|r-s|-1).
\eea
Inserting $c_{\alpha_1\beta_2}$ into \eqref{d'2} with $i=1, k=3$, we get
\bea
c_{\alpha_1\beta_3}=\delta(|r-s|-1),
\eea
Inserting $c_{\alpha_1\beta_3}$ into \eqref{d'2} with $i=1, k=4$, we get
\bea
c_{\alpha_1\beta_4}=\delta(|r-s|-1).
\eea
Repeating these steps, we can obtain $c_{\alpha_1\beta_k}=\delta(|r-s|-1)$ from $d'_{\alpha_1\beta_k}$ where $k=2,\dots,s$.

We then substitute $c_{\alpha_1\beta_1}$ into \eqref{d2} with $l=2,j=1$. Recalling that $d$ have the form \eqref{d0}, we get
\bea
c_{\alpha_2\beta_1}=\delta(|r-s|-1).
\eea
Substituting $c_{\alpha_2\beta_1}$ into \eqref{d'2} with $i=2,k=2$, we get
\bea
c_{\alpha_2\beta_2}=\delta(|r-s|-1).
\eea
Substituting $c_{\alpha_2\beta_2}$ into \eqref{d'2} with $i=2,k=3$, we get
\bea
c_{\alpha_2\beta_3}=\delta(|r-s|-1).
\eea
Repeating these steps, we solve that
$c_{\alpha_2\beta_k}=\delta(|r-s|-1)$ from $d'_{\alpha_2\beta_k}$ where $k=2,\dots, s$.

Following similar discussions and considering all the equations \eqref{d2} and \eqref{d'2}, we finally solve all the coefficients
\bea
c_{\alpha_i\beta_j}=\delta(|r-s|-1),~(i=1,\dots, r, j=1,\dots, s).
\eea
\item {\bf $c_{\alpha_i\alpha_j}$ and $c_{\beta_i\beta_j}$}

We consider $d'_{\alpha_i\alpha_j}$ in \eqref{A'(r,s)-1}. $d'_{\alpha_i\alpha_j}$ gets a $c_{\alpha_i\alpha_j}$ from each $\rho\in OP(\{\beta_1\}\bigcup\{\beta_2,\dots,\beta_s\})$. Thus we arrive at
\bea
sc_{\alpha_i\alpha_j}=s\delta(|s-r|-1).
\eea
Then $c_{\alpha_i\alpha_j}$ are solved as
\bea
c_{\alpha_i\alpha_j}=\delta(|s-r|-1).
\eea
If we consider $d_{\alpha_i\alpha_j}$ instead, we can solve $c_{\beta_i\beta_j}$  from \eqref{A(r,s)-1} in the same way. The solution is
\bea
c_{\beta_i\beta_j}=\delta(|s-r|-1).
\eea
\end{itemize}

To sum up, all the coefficients $c_{\alpha_i\alpha_j}$, $c_{\beta_i\beta_j}$ and $c_{\alpha_i\beta_j}$ in \eqref{gen-V(r,s)}
have the form $\delta(|s-r|-1)$. Considering on-shell condition and momentum conservation, the sum in \eqref{gen-V(r,s)} then give rise
\bea
{\cal V}^{(r,s)}=\left(1\over 2F^2\right)^{{r+s-1\over 2}}{1\over p_1^2}p_1^2\delta(|s-r|-1)=\left(1\over 2F^2\right)^{{r+s-1\over 2}}\delta(|s-r|-1).
\eea

\section{Conclusions}
In this paper, we proposed and proved the generalized $U(1)$-identity for tree-level off-shell currents in nonlinear sigma model.
When we take on-shell limit, this relation becomes the on-shell generalized $U(1)$ identity which is equivalent with KK relation. The $U(1)$-identity for off-shell currents proposed in \cite{Chen:2013fya} is a special case of the generalized $U(1)$-identity. There are several possible further extensions of this work, including the generalized off-shell BCJ relation, the loop-level extensions and the BCJ duality which implies the relations in nonlinear sigma model.

%%%%%%%%%%%%%%%%%%%%%%%%%%%%%%%%%%%%
\subsection*{Acknowledgements}
%%%%%%%%%%%%%%%%%%%%%%%%%%%%%%%%%%
Y. J. Du would like to acknowledge the EU programme Erasmus Mundus Action 2, Project 9  and
the International Postdoctoral Exchange Fellowship Program of China for supporting his postdoctoral research in Lund University (with Fudan University as the home university).
Y. J. Du's research is supported in parts by the NSF of China Grant No.11105118, China Postdoctoral Science Foundation No.2013M530175 and the Fundamental Research Funds for the Central Universities of Fudan University No.20520133169.
G. Chen, S. Y. Li and H. Q. Liu's  research  has been supported by the Fundamental Research Funds for the Central Universities under contract~020414340080, NSF of China Grant under contract~11405084, the Jiangsu Ministry of Science and Technology under contract~BK20131264 and by the Swedish Research Links programme of the Swedish Research Council (Vetenskapsradets generella villkor) under contract~348-2008-6049. We also thank Baoyi Chen, Edna Cheung, Yunxuan Li, Ruofei Xie, Yuan Xin for useful discussion.

\bibliographystyle{JHEP}
\bibliography{Refs}

\providecommand{\href}[2]{#2}\begingroup\raggedright\begin{thebibliography}{10}

\bibitem{Chen:2013fya}
G.~Chen and Y.-J. Du, {\it {Amplitude Relations in Non-linear Sigma Model}},
  {\em JHEP} {\bf 1401} (2014) 061,
  [\href{http://xxx.lanl.gov/abs/1311.1133}{{\tt arXiv:1311.1133}}].

\bibitem{Bern:2008qj}
Z.~Bern, J.~Carrasco, and H.~Johansson, {\it New relations for gauge-theory
  amplitudes},  {\em Phys. Rev. D} {\bf 78} (2008) 085011,
  [\href{http://xxx.lanl.gov/abs/0805.3993}{{\tt arXiv:0805.3993}}].

\bibitem{Kleiss:1988ne}
R.~Kleiss and H.~Kuijf, {\it Multi - gluon cross-sections and five jet
  production at hadron colliders},  {\em Nucl. Phys. B} {\bf 312} (1989)
  616--644.

\bibitem{BjerrumBohr:2009rd}
N.~Bjerrum-Bohr, P.~H. Damgaard, and P.~Vanhove, {\it {Minimal Basis for Gauge
  Theory Amplitudes}},  {\em Phys.Rev.Lett.} {\bf 103} (2009) 161602,
  [\href{http://xxx.lanl.gov/abs/0907.1425}{{\tt arXiv:0907.1425}}].

\bibitem{Stieberger:2009hq}
S.~Stieberger, {\it {Open $\&$ Closed vs. Pure Open String Disk Amplitudes}},
  \href{http://xxx.lanl.gov/abs/0907.2211}{{\tt arXiv:0907.2211}}.

\bibitem{DelDuca:1999rs}
V.~Del~Duca, L.~J. Dixon, and F.~Maltoni, {\it New color decompositions for
  gauge amplitudes at tree and loop level},  {\em Nucl. Phys. B} {\bf 571}
  (2000) 51--70, [\href{http://xxx.lanl.gov/abs/hep-ph/9910563}{{\tt
  hep-ph/9910563}}].

\bibitem{Feng:2010my}
B.~Feng, R.~Huang, and Y.~Jia, {\it {Gauge Amplitude Identities by On-shell
  Recursion Relation in S-matrix Program}},  {\em Phys.Lett.} {\bf B695} (2011)
  350--353, [\href{http://xxx.lanl.gov/abs/1004.3417}{{\tt arXiv:1004.3417}}].

\bibitem{Tye:2010kg}
H.~Tye and Y.~Zhang, {\it {Remarks on the identities of gluon tree
  amplitudes}},  {\em Phys.Rev.} {\bf D82} (2010) 087702,
  [\href{http://xxx.lanl.gov/abs/1007.0597}{{\tt arXiv:1007.0597}}].

\bibitem{Chen:2011jxa}
Y.-X. Chen, Y.-J. Du, and B.~Feng, {\it {A Proof of the Explicit Minimal-basis
  Expansion of Tree Amplitudes in Gauge Field Theory}},  {\em JHEP} {\bf 1102}
  (2011) 112, [\href{http://xxx.lanl.gov/abs/1101.0009}{{\tt
  arXiv:1101.0009}}].

\bibitem{Cachazo:2012uq}
F.~Cachazo, {\it {Fundamental BCJ Relation in N=4 SYM From The Connected
  Formulation}},  \href{http://xxx.lanl.gov/abs/1206.5970}{{\tt
  arXiv:1206.5970}}.

\bibitem{Bern:2010yg}
Z.~Bern, T.~Dennen, Y.-t. Huang, and M.~Kiermaier, {\it {Gravity as the Square
  of Gauge Theory}},  {\em Phys.Rev.} {\bf D82} (2010) 065003,
  [\href{http://xxx.lanl.gov/abs/1004.0693}{{\tt arXiv:1004.0693}}].

\bibitem{BjerrumBohr:2011xe}
N.~Bjerrum-Bohr, P.~Damgaard, H.~Johansson, and T.~Sondergaard, {\it
  {Monodromy--like Relations for Finite Loop Amplitudes}},  {\em JHEP} {\bf
  1105} (2011) 039, [\href{http://xxx.lanl.gov/abs/1103.6190}{{\tt
  arXiv:1103.6190}}].

\bibitem{Carrasco:2011mn}
J.~J. Carrasco and H.~Johansson, {\it {Five-Point Amplitudes in N=4
  Super-Yang-Mills Theory and N=8 Supergravity}},  {\em Phys.Rev.} {\bf D85}
  (2012) 025006, [\href{http://xxx.lanl.gov/abs/1106.4711}{{\tt
  arXiv:1106.4711}}].

\bibitem{Boels:2011tp}
R.~H. Boels and R.~S. Isermann, {\it {New relations for scattering amplitudes
  in Yang-Mills theory at loop level}},  {\em Phys.Rev.} {\bf D85} (2012)
  021701, [\href{http://xxx.lanl.gov/abs/1109.5888}{{\tt arXiv:1109.5888}}].

\bibitem{Boels:2011mn}
R.~H. Boels and R.~S. Isermann, {\it {Yang-Mills amplitude relations at loop
  level from non-adjacent BCFW shifts}},  {\em JHEP} {\bf 1203} (2012) 051,
  [\href{http://xxx.lanl.gov/abs/1110.4462}{{\tt arXiv:1110.4462}}].

\bibitem{Bern:2012uf}
Z.~Bern, J.~Carrasco, L.~Dixon, H.~Johansson, and R.~Roiban, {\it {Simplifying
  Multiloop Integrands and Ultraviolet Divergences of Gauge Theory and Gravity
  Amplitudes}},  {\em Phys.Rev.} {\bf D85} (2012) 105014,
  [\href{http://xxx.lanl.gov/abs/1201.5366}{{\tt arXiv:1201.5366}}].

\bibitem{Carrasco:2012ca}
J.~J.~M. Carrasco, M.~Chiodaroli, M.~G¨¹naydin, and R.~Roiban, {\it {One-loop
  four-point amplitudes in pure and matter-coupled N $\leq$ 4 supergravity}},
  {\em JHEP} {\bf 1303} (2013) 056,
  [\href{http://xxx.lanl.gov/abs/1212.1146}{{\tt arXiv:1212.1146}}].

\bibitem{Du:2012mt}
Y.-J. Du and H.~Luo, {\it {On General BCJ Relation at One-loop Level in
  Yang-Mills Theory}},  {\em JHEP} {\bf 1301} (2013) 129,
  [\href{http://xxx.lanl.gov/abs/1207.4549}{{\tt arXiv:1207.4549}}].

\bibitem{Oxburgh:2012zr}
S.~Oxburgh and C.~White, {\it {BCJ duality and the double copy in the soft
  limit}},  {\em JHEP} {\bf 1302} (2013) 127,
  [\href{http://xxx.lanl.gov/abs/1210.1110}{{\tt arXiv:1210.1110}}].

\bibitem{Saotome:2012vy}
R.~Saotome and R.~Akhoury, {\it {Relationship Between Gravity and Gauge
  Scattering in the High Energy Limit}},  {\em JHEP} {\bf 1301} (2013) 123,
  [\href{http://xxx.lanl.gov/abs/1210.8111}{{\tt arXiv:1210.8111}}].

\bibitem{Boels:2012ew}
R.~H. Boels, B.~A. Kniehl, O.~V. Tarasov, and G.~Yang, {\it {Color-kinematic
  Duality for Form Factors}},  {\em JHEP} {\bf 1302} (2013) 063,
  [\href{http://xxx.lanl.gov/abs/1211.7028}{{\tt arXiv:1211.7028}}].

\bibitem{Boels:2013bi}
R.~H. Boels, R.~S. Isermann, R.~Monteiro, and D.~O'Connell, {\it
  {Colour-Kinematics Duality for One-Loop Rational Amplitudes}},  {\em JHEP}
  {\bf 1304} (2013) 107, [\href{http://xxx.lanl.gov/abs/1301.4165}{{\tt
  arXiv:1301.4165}}].

\bibitem{Bjerrum-Bohr:2013iza}
N.~E.~J. Bjerrum-Bohr, T.~Dennen, R.~Monteiro, and D.~O'Connell, {\it
  {Integrand Oxidation and One-Loop Colour-Dual Numerators in N=4 Gauge
  Theory}},  {\em JHEP} {\bf 1307} (2013) 092,
  [\href{http://xxx.lanl.gov/abs/1303.2913}{{\tt arXiv:1303.2913}}].

\bibitem{Bern:2013yya}
Z.~Bern, S.~Davies, T.~Dennen, Y.-t. Huang, and J.~Nohle, {\it
  {Color-Kinematics Duality for Pure Yang-Mills and Gravity at One and Two
  Loops}},  \href{http://xxx.lanl.gov/abs/1303.6605}{{\tt arXiv:1303.6605}}.

\bibitem{Nohle:2013bfa}
J.~Nohle, {\it {Color-Kinematics Duality in One-Loop Four-Gluon Amplitudes with
  Matter}},  {\em Phys.Rev.} {\bf D90} (2014) 025020,
  [\href{http://xxx.lanl.gov/abs/1309.7416}{{\tt arXiv:1309.7416}}].

\bibitem{Mafra:2011kj}
C.~R. Mafra, O.~Schlotterer, and S.~Stieberger, {\it {Explicit BCJ Numerators
  from Pure Spinors}},  {\em JHEP} {\bf 1107} (2011) 092,
  [\href{http://xxx.lanl.gov/abs/1104.5224}{{\tt arXiv:1104.5224}}].

\bibitem{Monteiro:2011pc}
R.~Monteiro and D.~O'Connell, {\it {The Kinematic Algebra From the Self-Dual
  Sector}},  {\em JHEP} {\bf 1107} (2011) 007,
  [\href{http://xxx.lanl.gov/abs/1105.2565}{{\tt arXiv:1105.2565}}].

\bibitem{BjerrumBohr:2012mg}
N.~Bjerrum-Bohr, P.~H. Damgaard, R.~Monteiro, and D.~O'Connell, {\it {Algebras
  for Amplitudes}},  {\em JHEP} {\bf 1206} (2012) 061,
  [\href{http://xxx.lanl.gov/abs/1203.0944}{{\tt arXiv:1203.0944}}].

\bibitem{Fu:2012uy}
C.-H. Fu, Y.-J. Du, and B.~Feng, {\it {An algebraic approach to BCJ
  numerators}},  {\em JHEP} {\bf 1303} (2013) 050,
  [\href{http://xxx.lanl.gov/abs/1212.6168}{{\tt arXiv:1212.6168}}].

\bibitem{Monteiro:2013rya}
R.~Monteiro and D.~O'Connell, {\it {The Kinematic Algebras from the Scattering
  Equations}},  {\em JHEP} {\bf 1403} (2014) 110,
  [\href{http://xxx.lanl.gov/abs/1311.1151}{{\tt arXiv:1311.1151}}].

\bibitem{Broedel:2011pd}
J.~Broedel and J.~J.~M. Carrasco, {\it {Virtuous Trees at Five and Six Points
  for Yang-Mills and Gravity}},  {\em Phys.Rev.} {\bf D84} (2011) 085009,
  [\href{http://xxx.lanl.gov/abs/1107.4802}{{\tt arXiv:1107.4802}}].

\bibitem{Fu:2014pya}
C.-H. Fu, Y.-J. Du, and B.~Feng, {\it {Note on symmetric BCJ numerator}},  {\em
  JHEP} {\bf 1408} (2014) 098, [\href{http://xxx.lanl.gov/abs/1403.6262}{{\tt
  arXiv:1403.6262}}].

\bibitem{Naculich:2014rta}
S.~G. Naculich, {\it {Scattering equations and virtuous kinematic numerators
  and dual-trace functions}},  {\em JHEP} {\bf 1407} (2014) 143,
  [\href{http://xxx.lanl.gov/abs/1404.7141}{{\tt arXiv:1404.7141}}].

\bibitem{Cachazo:2013gna}
F.~Cachazo, S.~He, and E.~Y. Yuan, {\it {Scattering Equations and KLT
  Orthogonality}},  {\em Phys.Rev.} {\bf D90} (2014) 065001,
  [\href{http://xxx.lanl.gov/abs/1306.6575}{{\tt arXiv:1306.6575}}].

\bibitem{Cachazo:2013hca}
F.~Cachazo, S.~He, and E.~Y. Yuan, {\it {Scattering of Massless Particles in
  Arbitrary Dimensions}},  {\em Phys.Rev.Lett.} {\bf 113} (2014), no.~17
  171601, [\href{http://xxx.lanl.gov/abs/1307.2199}{{\tt arXiv:1307.2199}}].

\bibitem{Cachazo:2013iea}
F.~Cachazo, S.~He, and E.~Y. Yuan, {\it {Scattering of Massless Particles:
  Scalars, Gluons and Gravitons}},  {\em JHEP} {\bf 1407} (2014) 033,
  [\href{http://xxx.lanl.gov/abs/1309.0885}{{\tt arXiv:1309.0885}}].

\bibitem{Naculich:2014naa}
S.~G. Naculich, {\it {Scattering equations and BCJ relations for gauge and
  gravitational amplitudes with massive scalar particles}},  {\em JHEP} {\bf
  1409} (2014) 029, [\href{http://xxx.lanl.gov/abs/1407.7836}{{\tt
  arXiv:1407.7836}}].

\bibitem{Bern:2011ia}
Z.~Bern and T.~Dennen, {\it {A Color Dual Form for Gauge-Theory Amplitudes}},
  {\em Phys.Rev.Lett.} {\bf 107} (2011) 081601,
  [\href{http://xxx.lanl.gov/abs/1103.0312}{{\tt arXiv:1103.0312}}].

\bibitem{Du:2013sha}
Y.-J. Du, B.~Feng, and C.-H. Fu, {\it {The Construction of Dual-trace Factor in
  Yang-Mills Theory}},  {\em JHEP} {\bf 1307} (2013) 057,
  [\href{http://xxx.lanl.gov/abs/1304.2978}{{\tt arXiv:1304.2978}}].

\bibitem{Fu:2013qna}
C.-H. Fu, Y.-J. Du, and B.~Feng, {\it {Note on Construction of Dual-trace
  Factor in Yang-Mills Theory}},  {\em JHEP} {\bf 1310} (2013) 069,
  [\href{http://xxx.lanl.gov/abs/1305.2996}{{\tt arXiv:1305.2996}}].

\bibitem{Du:2014uua}
Y.-J. Du, B.~Feng, and C.-H. Fu, {\it {Dual-color decompositions at one-loop
  level in Yang-Mills theory}},  {\em JHEP} {\bf 1406} (2014) 157,
  [\href{http://xxx.lanl.gov/abs/1402.6805}{{\tt arXiv:1402.6805}}].

\bibitem{Bargheer:2012gv}
T.~Bargheer, S.~He, and T.~McLoughlin, {\it {New Relations for
  Three-Dimensional Supersymmetric Scattering Amplitudes}},  {\em
  Phys.Rev.Lett.} {\bf 108} (2012) 231601,
  [\href{http://xxx.lanl.gov/abs/1203.0562}{{\tt arXiv:1203.0562}}].

\bibitem{Ma:2011um}
Q.~Ma, Y.-J. Du, and Y.-X. Chen, {\it {On Primary Relations at Tree-level in
  String Theory and Field Theory}},  {\em JHEP} {\bf 1202} (2012) 061,
  [\href{http://xxx.lanl.gov/abs/1109.0685}{{\tt arXiv:1109.0685}}].

\bibitem{Du:2011js}
Y.-J. Du, B.~Feng, and C.-H. Fu, {\it {BCJ Relation of Color Scalar Theory and
  KLT Relation of Gauge Theory}},  {\em JHEP} {\bf 1108} (2011) 129,
  [\href{http://xxx.lanl.gov/abs/1105.3503}{{\tt arXiv:1105.3503}}].

\bibitem{Kampf:2012fn}
K.~Kampf, J.~Novotny, and J.~Trnka, {\it {Recursion Relations for Tree-level
  Amplitudes in the SU(N) Non-linear Sigma Model}},  {\em Phys.Rev.} {\bf D87}
  (2013) 081701, [\href{http://xxx.lanl.gov/abs/1212.5224}{{\tt
  arXiv:1212.5224}}].

\bibitem{Kampf:2013vha}
K.~Kampf, J.~Novotny, and J.~Trnka, {\it {Tree-level Amplitudes in the
  Nonlinear Sigma Model}},  {\em JHEP} {\bf 1305} (2013) 032,
  [\href{http://xxx.lanl.gov/abs/1304.3048}{{\tt arXiv:1304.3048}}].

\bibitem{Berends:1988zn}
F.~A. Berends and W.~Giele, {\it {Multiple Soft Gluon Radiation in Parton
  Processes}},  {\em Nucl.Phys.} {\bf B313} (1989) 595.

\end{thebibliography}\endgroup
\end{document}